\documentclass[a4paper,11pt]{article}
\usepackage{pos}
 \usepackage{graphicx}
\usepackage{graphicx}
\usepackage[utf8]{inputenc}
\usepackage[english]{babel}
\usepackage{xcolor}

\usepackage[utf8]{inputenc}
\usepackage[english]{babel}
\usepackage{xcolor}

\date{``Korfu''   , September , 2021}

\title{Nanometer 
  Size Dirty Dark Matter Pearls, $e^-$-signal, IMP or SIDM, not WIMP
\footnote{IMP=`` Interacting Massive Particles ''.

SIDM = ``Self interacting Dark Matter''.}}

\author*[a]{Holger Bech Nielsen}
\author[a,b]{Colin D. Froggatt}

\affiliation[a]{Niels Bohr Institute,\\
  Blegdamsvej 15  - 21, Copenhagen, Denmark}

\affiliation[b]{Physics and Astronomy, University of Glasgow,\\
  Glasgow, Scotland}

\emailAdd{hbech@nbi.dk}
\emailAdd{Colin.Froggatt@Glasgow.ac.uk}

\abstract{Through several articles we have developed a model for dark matter as
  consisting of bubbles of a new (speculated) type of vacuum, starting from
  cm-sized pearls or balls down to atomic size ones and now we believe they
  have nanometer sizes. In the latest development of our model we have the
  bubbles of the new vacuum imbedded in dust grains very similar to the
  grains present in interstellar and intergalactic space anyway, although the
  presence of the bubble with a very large homolumo gap in its single electron
  spectrum influences the dust grain material so as to become denser and harder.

  We have earlier explained how our dark matter particles get stopped in the
  shielding, so that normally expected nucleonic collisions are not observable.
  The signal of the dark matter in the underground experiments 
  rather becomes decays of excited
  particles actually with the energy of the homolumo gap, which is also 
  equal to the photon energy of the 
  X-ray line presumably observed astronomically from galaxy clusters etc.
  
  A new calculation here is a fitting of the 
  velocity dependence of the dark matter self-interaction as estimated by 
  Correa \cite{CAC}, using deviations from the only
  gravitationally interacting dark matter in dwarf galaxies.

  Let us stress that apart from the speculated new vacuum we have
  no new physics,
  and if the couplings in the Standard Model were adjusted to make degenerate
  vacua as speculated according to our Multiple Point 
  Principle (MPP) we would only need the Standard Model, so dark matter would
  {\em not} require new physics.}

\FullConference{%
  Corfu Summer Institute 2021 "School and Workshops on Elementary Particle Physics and Gravity"\\
  29 August - 9 October 2021\\
  Corfu, Greece
}

\begin{document}
\maketitle

\titlepage

\section{Introduction}
One of the most mysterious facts about the dark matter is that the
DAMA-LIBRA experiment, which only recognizes dark matter from its
annually varying impact rate, sees the dark matter, while seemingly very
similar experiments using xenon and able to distinguish, if the dark matter
particle has hit a nucleus or say an electron, do not see dark matter.
We take this mystery to point towards models of the type of
Khlopov \cite{Khlopov,Khlopov1, Khlopov3, Khlopov5} , wherein
the  dark matter particle is larger and more composite, as e.g. Khlopov's
model with an a priori doubly negatively charged object having been
neutralized by a He-nucleus.

In fact we have proposed for some time dark matter model(s) based on the idea
of having several vacua with the same energy density  \cite{extension, Dark1, Dark2, Tunguska, supernova, Corfu2017, Corfu2019, theline, Bled20, Bled21}.
Then the dark matter particle
is a bubble or pearl of one of the new vacua filled with some ordinary matter
- e.g. diamond - under a very high pressure caused by the surface tension of
the surface - the domain wall - separating the two phases of vacuum. But
recently we got to expecting these bubbles to be surrounded by an essentially
normal dust grain, of which there are so many anyway in the interstellar or
intergalactic medium. This dust material may though be made harder and more
dense by effects of the homolumo gap in the pearl influencing the material of
the dust. 

The crux of the matter for our model is now that the dark matter pearls
are so large that they interact significantly with each other and with
ordinary matter. In fact especially Correa \cite{CAC} found phenomenological
evidence for an even velocity dependent cross section in as far as the ratio
$\frac{\sigma}{M}$ of the cross section $\sigma$ to the mass $M$, which  is
the only
quantity you can extract when the observation is indirect in the sense
that you only can measure the dark matter via its influence on the ordinary
matter, stars say, turned out to fall with velocity. It is shown in 
Figure
\ref{Correa} how this ratio falls off with collision velocity $v$.

A rather recent calculation consists in seeking to understand the velocity
dependence of the ratio $\frac{\sigma}{M}$ in terms of our model
of a highly dense pearl surrounded by a rather usual dust grain only made
somewhat harder and more condensed by the homolumo gap effect on the
dust grain.

When the dark matter particles reach the Earth we imagine that the
dust grain around the genuine new vacuum bubble gets boiled off, presumably
already in the atmosphere. At the same time the pearl gets excited
and some electrons in the interior get excited across the homolumo gap.
This leads some electrons first losing energy to about 
the lumo-state energy (just at the bottom of the unfilled states) and, if some
excitations can survive so long that the excited pearl can reach down to the
experiments looking for dark matter underground, the emission of photons or
electrons with energies given by the homolumo gap may be detected. 
So our model predicts
that the underground observations should mainly be electron events
just with energy equal to homolumo gap size 3.5 keV - taken to be so as to
identify such decays also with the source of the 3.5 keV X-ray radiation
observed by satellites \cite{Bulbul, Boyarsky, Boyarsky2, Bhargava, Sicilian, Foster}.  
It is indeed remarkable that the average of
the energy per event, both in the DAMA-LIBRA experiment \cite{DAMA1, DAMA2}
and in the
recently observed mysterious excess of events seemingly coming from electron
collisions observed with low statistics by Xenon1T \cite{Xenon1Texcess}, 
is close to the value 3.5 keV!

This difference of our model from the most popular type
of model, the WIMP type of models, can be claimed by saying:
Our model could be called IMP or SIDM but not WIMP, where
%
%
  IMP =``Interacting Massive Particles'' and
SIDM = ``Self-Interacting Dark Matter''.


  A bit similar to Khlopov's model of double negatively charged particles
  leading to zero charged He that can be stopped, our model assumes:
  \begin{itemize}
  \item {\em The dark matter particles interact so strongly as to
    get stopped in the earth shielding before reaching the
    direct dark matter
    searches like DAMA, LUX, Xenon1T,...}
  \item {\em They can get excited and emit energy as electrons or
    photons, say hours or even years  after hitting the Earth.}. 
  \end{itemize}
  Thus DAMA-LIBRA may see them as dark matter, but experiments needing
  recoil nuclei will not accept them as dark matter.
\section{Pearl}
  \subsection{Dark Nanometer  Size Pearls, Electronic 3.5 keV Signal,
    IMP or SIDM, not WIMP}
  
  
  \begin{figure}
    \includegraphics{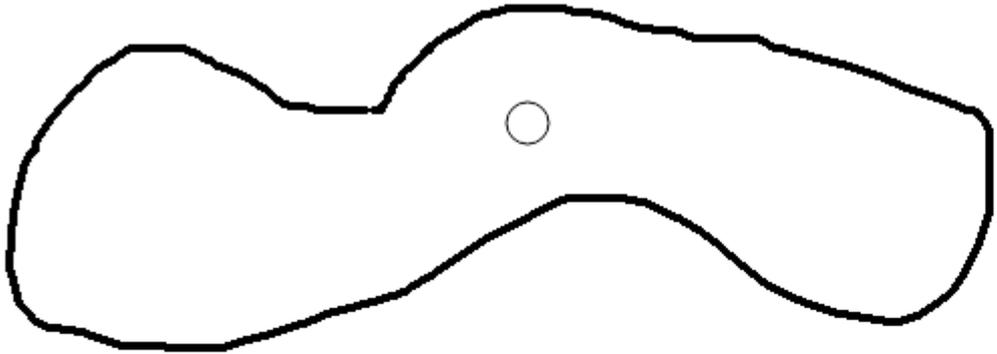}
    \caption{\label{crude}
      Simplified crude picture of our dark matter particle
      as a small accurately spherical bubble of new vacuum with some ordinary
      matter inside, say carbon, under high pressure. Around it is a more
      irregular grain of dust having been collected by the little bubble,
      much like an ordinary grain of dust in the intergalactic or interstellar
      space.}
    \end{figure}
 
   Figure \ref{crude} illustrates a fraction of a micrometer size 
    small dust grain with an about a nanometer big new vacuum
    bubble inside. This
    complicated/macroscopic object 
    corresponds to a dark matter particle in our model.
    There are about $10^{12}/12 \approx 10^{11}$ carbon ``atoms'' 
     inside the bubble.

  \begin{itemize}
  \item In the middle is a spherical bubble of radius
    \begin{eqnarray}
      R \approx r_{cloud \; 3.3 MeV}&\approx &
      10^{-9}m
    \end{eqnarray}
     Here $r_{cloud \; 3.3 MeV}$ denotes the radius where the electron potential
    is 3.3 MeV, which is identified with the Fermi energy $E_f$ of the
    electrons in the bulk of the pearl - i.e. inside the radius $R$. We
    estimated the value $E_f = 3.3$ MeV in previous papers \cite{Corfu2019, theline, Bled20} by fitting the
    overall rate of the intensity of the 3.5 keV line emitted by galactic
    clusters and the very frequency 3.5 keV of the radiation in our model.
  \item The radius $r_{cloud \; 3.5 keV}$ is where the electron potential
    is 3.5 keV.  By our story of the ``homolumo gap'':  the
    electron density crudely goes to zero at this radius. (It falls
    a lot in the range between $r_{cloud \; 3.3 MeV} $ and
    $r_{cloud \; 3.5 keV}$). The distance between the two different
    radii $R=r_{cloud \; 3.3 MeV}$ and $r_{cloud \; 3.5 keV}$ is only
    about $10^{-12}m$, so with nanometer size pearls the essentially
    purely electron region between the two is only a rather thin layer.
    \end{itemize}


  \subsection{The electron density and potential in the pearls}
\begin{figure}
  \includegraphics[scale=0.8]{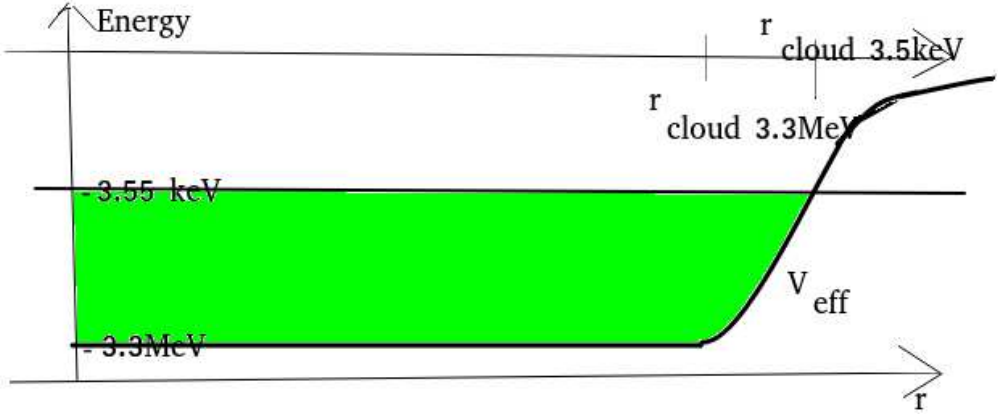}
  \caption{The green filled out region symbolizes the Fermi-sea
    of electrons. They have energies above the effective potential,
    which is the lower end of this `green'' region. On the left axis is the
    electron energy. The numbers $r_{cloud \; 3.5 keV} $ and $r_{cloud \; 3.3 MeV}$
    give the distances from the center to the points where the potential
    numerically has the values in the index.} 
 \label{density}
  \end{figure}

  \begin{itemize}
  \item Due to an effect we call the {\em homolumo-gap effect} \cite{Corfu2017, jahn}, the nuclei in the
    bubble region and the electrons themselves become arranged in such a way
    as to prevent there from being any levels in an interval of width
    3.5 keV. So outside the distance $r_{3.5 keV} = r_{cloud \; 3.5 keV}$ from
    the center of the pearl at which the Coulomb potential is $\sim 3.5 keV$
    deep there are essentially ($\sim$  in the Thomas-Fermi approximation) no
    more electrons in the pearl-object, as illustrated in Figure \ref{density} .
  \item The radius $r_{3.3 MeV}= r_{cloud \; 3.3 MeV}$ at which the potential
    felt by an electron is $3.3 MeV$ deep, is  supposed to be just the radius
    to which the many nuclei
    replacing the only one nucleus in ordinary atoms reach out. So inside
    it the potential is much more flat.
  \end{itemize}

  \begin{itemize}
  \item The energy difference between the zero energy line and the effective
    Fermi surface above which there are no more electrons is of order of 
    3.5 keV, the value so crucial to our work.
  \item Since in the Thomas Fermi approximation there are no electrons outside
    roughly the
    radius $r_{3.5 keV}=r_{cloud \; 3.5 keV}$, this radius will give the maximal
    self-interaction cross section, even for very low velocity 
    $\sigma_{v\rightarrow 0}$, were it
    not for the dust grain supposedly having built up around the pearl. We
    shall discuss the dust extension more below. 
    \end{itemize}



\begin{figure}
\includegraphics{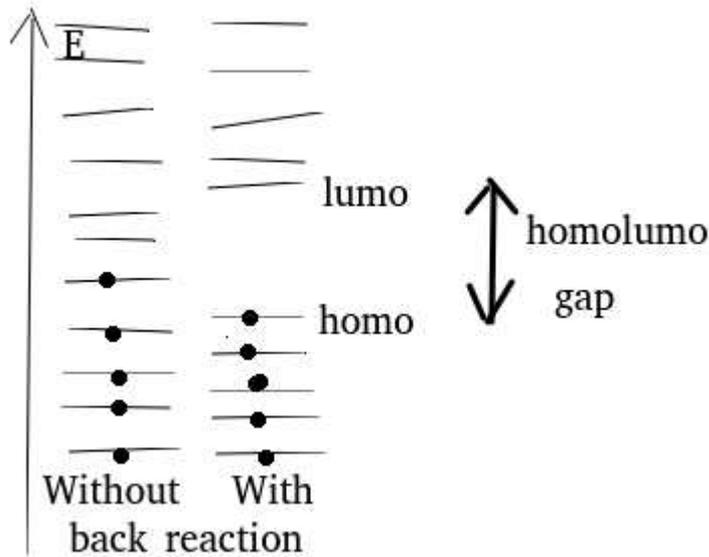}
\caption{The electron energy spectrum in the left
  column is supposed to be what one would get without {\em back reaction},
  meaning ignoring that the electrons act back on the nuclei and other
  electrons around, so as to put them in a way influenced by the electrons
  in the electron eigenstates. The spectrum to the right is supposed to
  be modified relative to the one to the left by such a back reaction.
  The energy of the total system can be lowered , if the nuclei and other
  electrons adjust so as to lower the filled levels.} 
\label{homolumo}
\end{figure}

        \subsection{The homolumo-gap effect.}

        Think first of the spectrum of energy levels for the electrons
        by assuming at the start the positions or distributions of the charged
        particles in the piece of material studied, e.g. one of our pearls,
        as fixed.

        Then the ground state is just a state built e.g. as a Slater
        determinant for the electrons being in the lowest single electron
        states,
        so many as are needed to have the right number of electrons.

        But now if the charged particles can be moved due to their
        interactions, the
        ground state energy could be lowered by moving them so that the
        energy of the filled
        electron state levels get lowered.

        {\em So we expect introducing such a ``back reaction'' will lower
          the filled states.}


        When the filled levels get moved downwards,
         then the homo= ``{\bf h}ighest {\bf o}ccupied {\bf m}olecular
        {\bf o}rbit'' level will be lowered and its distance to the
        lumo=``{\bf l}owest {\bf u}noccupied {\bf m}olecular {\bf o}rbit''
        increased by the effect of the back  reaction, and so an
        exceptionally large region, the ``homolumo-gap'' with no single
        particle electron levels,
        will appear on the energy axis, as illustrated in Figure \ref{homolumo}.


        {\em We believe we can estimate the homolumo-gap $E_H$.}
        
        Using the Thomas Fermi approximation - or crudely just some dimensional
        argument, where the fine structure constant has dimension of velocity -
        we  calculated the homolumo gap in highly compressed ordinary
        matter, for relativistic electrons \cite{theline}:
        \begin{eqnarray}
          E_H &\sim& (\frac{\alpha}{c})^{3/2}\sqrt{2}p_f\\
          \hbox{where } \quad p_f &=& \hbox{Fermi momentum} \\
          \frac{\alpha}{c}&=& \frac{1}{137.03...}
        \end{eqnarray}
        (the $\sqrt{2}$ comes from our Thomas Fermi calculation).

        It is by requiring this homolumo-gap to be the 3.5 keV energy per photon
        of the X-ray radiation mysteriously observed by satellites
        astronomically from clusters of galaxies, Andromeda and the Milky Way 
        Center \cite{Bulbul, Boyarsky, Boyarsky2, Bhargava, Sicilian, Foster}
        that we estimate the Fermi-energy to be $E_f \approx p_f = 3.3$ MeV 
        in the interior bulk of the pearl.
        
       
       \section{The Dirty Pearls}
       
\begin{figure}!
       	\includegraphics
       	{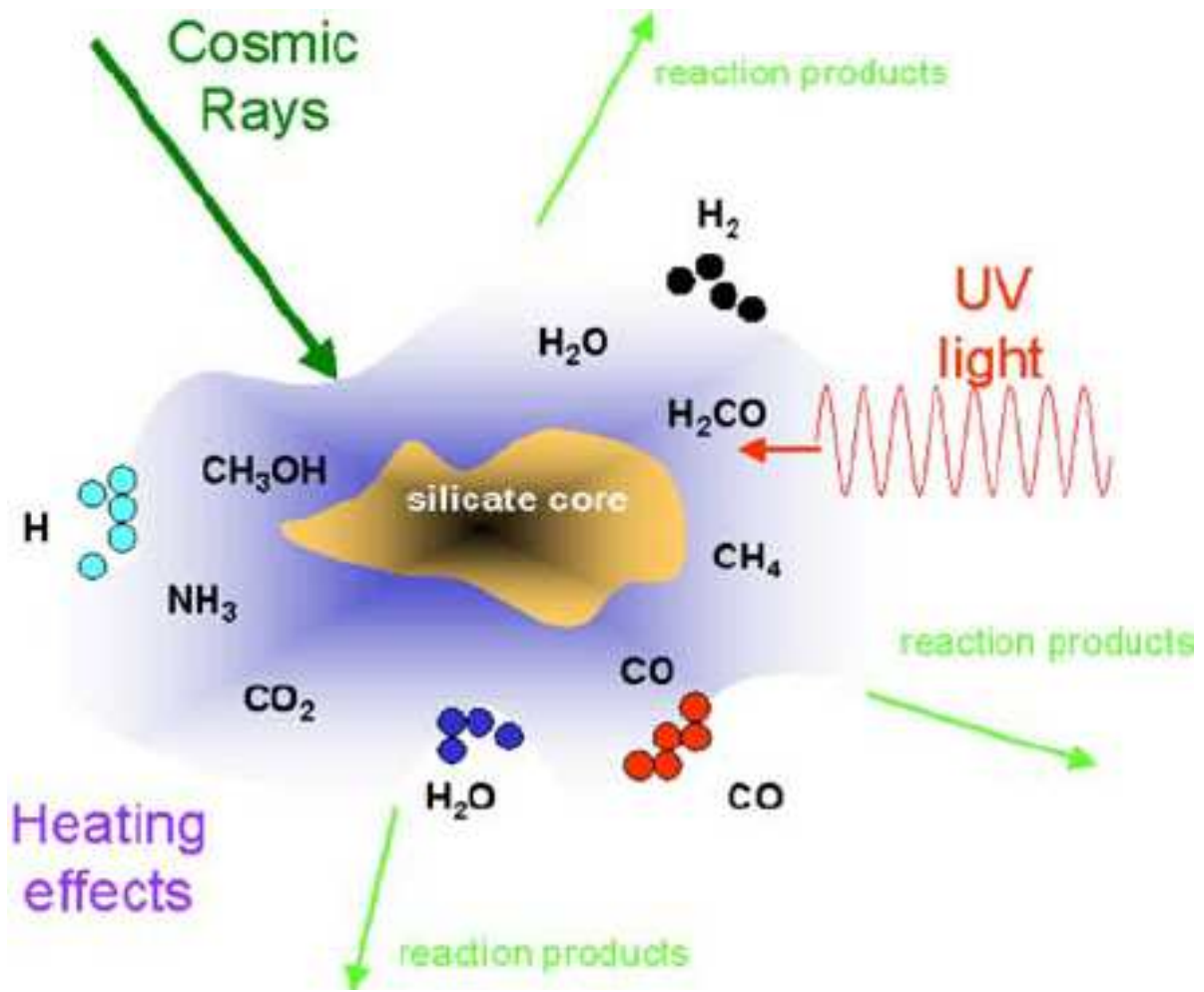}
       	\caption{\label{df} Model of Dust Particle}
       \end{figure}

       The most important and heaviest parts of the dark matter particles are
       the small regions of a new vacuum just described with its surrounding
       range of electrons which, because of the high pressure in the degenerate
       collection of electrons, tend to be somewhat pressed out compared to the
       nuclei which, because they are not degenerate, are easier to be kept
       inside. However, these pearls survive in the intergalactic or
       interstellar space for more than 13 milliard years, and so we expect them
       to become dirty. In fact there exist dust grains in space, typically
       of the order of $0.1\mu m$ in size and consisting of silicates or 
       chemical compositions with carbon. A typical model picture of
       such a dust grain in the outer space is given in Figure \ref{df}. 
       
\begin{figure}
       	\includegraphics{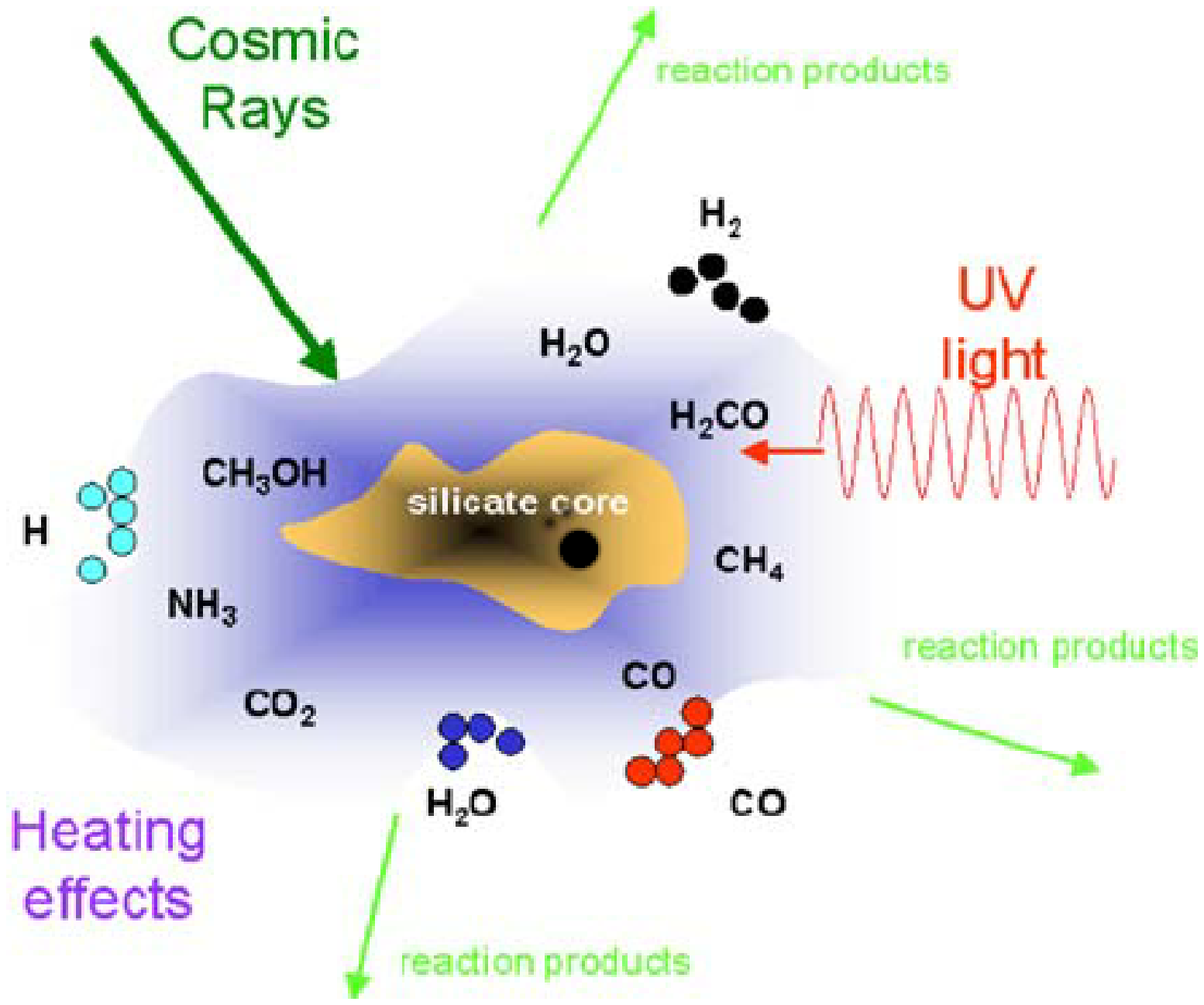}
       	\caption{\label{df2} The little dot inserted in the foregoing figure here
       		symbolizes
       		the bubble of new vacuum. It is supposedly much heavier than the
       		rest of the dust grain.}
       \end{figure}

       It is indeed very likely that such a dust grain would collect
       in many cases on top of one of our dark matter particles, 
       which is in itself
       very much like a seed atom. We may illustrate that by drawing our little
       pearl with its new vacuum as a ball inside the usual dust grain model
       picture \ref{df2}.
       
       \subsection{Influence of Homolumo gap on the Dust Matter}
       \label{IHGDM}
       In this subsection we want to speculate that the adjustment of the
       degrees of freedom mainly in the pearl or bubble of new vacuum 
       will influence the properties of the dust grain built up
       around this little heavy pearl - we shall return in section
       \ref{ehg} to this subject. The argument suggesting that
       a rather strong effect of the homolumo gap is expected is the following:
       we might consider the whole object - dust grain plus the
       bubble in the middle - as one could say a molecule, so that
       the single electron spectrum {\em for the whole
       	dark matter particle with both dust and bubble} should have
       the homolumo gap 3.5 keV. Now this gap size 3.5 keV is for
       usual chemistry a very high number, only reached as a binding energy
       for the very tightest bound electrons. Thus if the wave functions filled
       in the whole dark matter object should have the binding energy of the
       order of 3.5 keV, it would either be needed 
       that an electron wave function would
       have so much overlap with the interior pearl region that this overlap
       could quite dominate the binding energy of the whole wave function, or
       the part of the wave function in the region outside in the dust grain
       should concentrate around nuclei very strongly. Essentially in this latter
       possibility only the most strongly bound electronic orbits in the dust
       region should be filled.
       
       To get an idea about, what we should expect the filled wave functions
       to look like in the outside region in the dust grain, let us think of
       constructing these wave functions in this dust grain region by expanding
       them as macroscopically slowly varying  superpositions of one orbit around
       the atoms in the outside. This is very close to what one does by
       using Bloch wave functions in the tight binding approximation in solid
       state physics.
       
       In fact, if we take a wave function ansatz to be composed from only
       one specific atomic level with binding energy $E_l$ as a from atom to
       atom smoothly varying function $\psi(\vec{x})$ of the atom positions,
       we can in the macroscopic approximation, that there are many atoms
       inside the length scale we are interested in, describe the electron as
       behaving freely except that it feels a constant negative potential of
       $E_l$ in the whole dust grain region. Only in the very small sphere, where
       the bubble sits, there is a very strong and quite different effective
       potential. But now we must - assuming that the
       homolumo gap for the whole system shall be of the large value 3.5 keV -
       believe that the system has adjusted, so that all the filled states have
       negative energy of the order of 3.5 keV. This means that considering
       a wave function ansatz constructed from the upper levels in the atoms in
       the dust grain region - thus having only relatively small potential
       $E_l$ - it will be strongly / exponentially concentrated around the
       bubble (with its high binding). Only if the atomic level from which
       we construct the wave function ansatz has a numerically large
       binding $E_l$, preferably -3.5 keV, will the ansatz wave function 
       be able to extend far out in the dust grain.
       
       Thus it appears that looking at the outskirts of the dust grain
       the electrons in the filled states reaching out there must be
       essentially all along in the most strongly bound atomic orbits for the
       atoms in the dirt. This means that this dirt-material should be
       imagined to be significantly different in internal structure from
       the dirt in a dust grain without our bubble pearl inside it.
       
       Actually we expect the dust grain with a bubble inside to have
       the electrons only in the lowest atomic orbits bound by orders of
       keV rather than as usual eV. This might even mean that these atoms
       become positive ions in as far as there will be too few
       deeply bound atomic orbits. However, presumably the density of the nuclei
       can adjust so that the total charge density can be zero anyway.
       
       Next one may then wonder how this by the homolumo gap modified
       structure of the atoms would influence the properties of the
       material. Presumably the geometrically smaller size of the
       deeper bound orbits would make the density of the dust
       material higher. The atoms can come closer. Also the scale of the
       binding energy of the atoms to each other would likely be scaled up
       and the material would be harder. 
      

        

\begin{figure}!
      
      \includegraphics{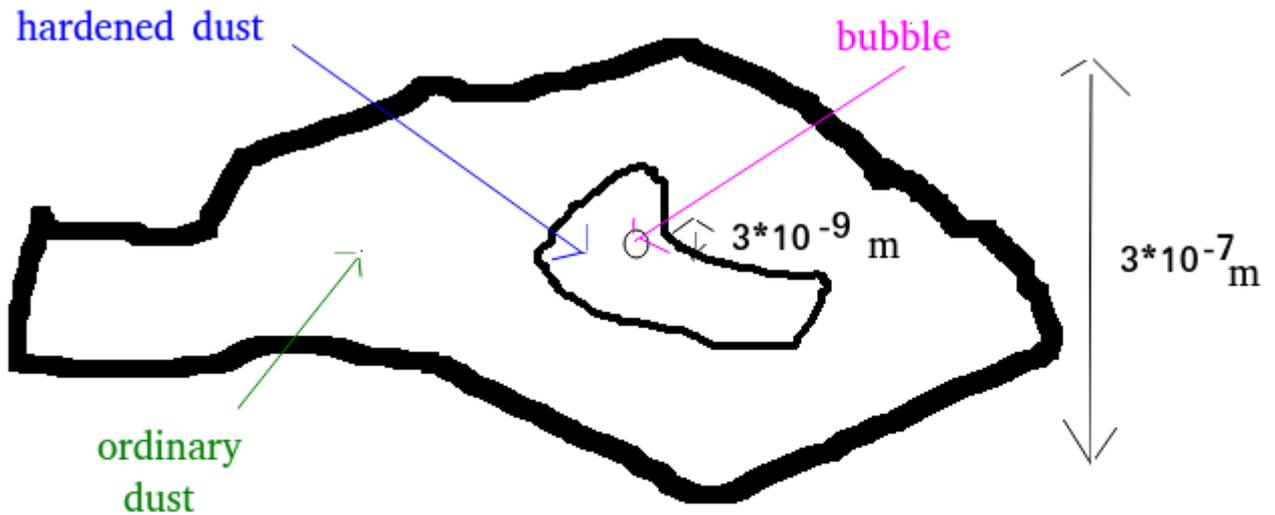}
      \caption{The crude dimensions of our dark matter model particle including
        its dust grain, which is typically irregular. The much smaller
        genuine bubble of new vacuum is, however, very accurately spherical,
        because it is so as to minimize the skin or surface area
        for the essentially
        given volume of highly compressed ordinary material on the background
        of the second vacuum. In the dust material but near to the pearl
        we expect the stuff to be hardened due to an effect from the homolumo
        gap, which inside the bubble has for ordinary matter an unusually high
        value $3.5$ keV.}
    \label{dimensions}
\end{figure}

 \begin{figure}

    \includegraphics[scale=0.9]{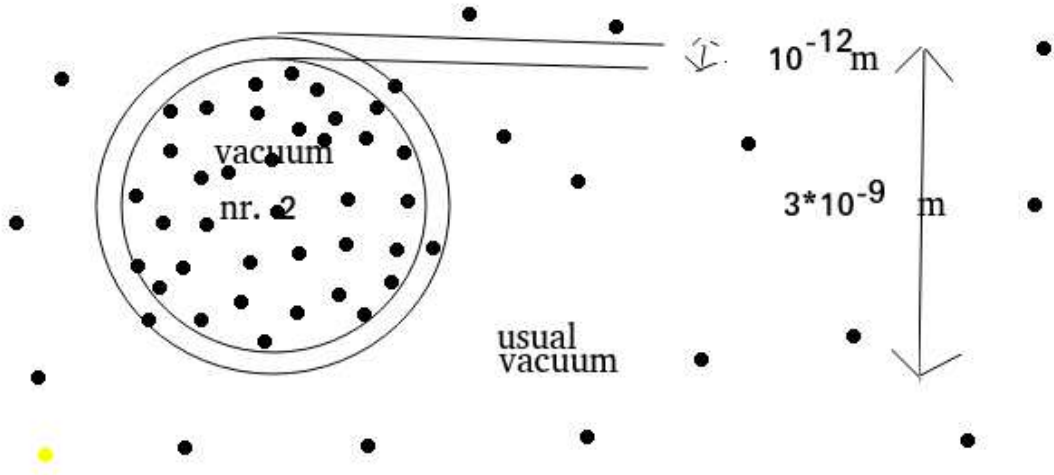}
    
    \caption{Illustration of the bubble of new vacuum as imbedded into
      a dust grain, which is not fully drawn. The size of the
      bubble is of the order of a few nanometers surrounded by a
      much thinner layer of electrons having been pushed a bit,
      about $10^{-12}m$, outside the domain wall separating the
      two vacuum phases. Because of the high pressure from this domain
      wall (=skin), the density of atomic nuclei inside the skin is
    expected to
    be larger than outside, even though the dust material just
    outside should have been hardened by an effect of the homolumo gap
    (see subsection \ref{IHGDM}). The density of nuclei shown on the figure
    here, both inside and outside, is lower compared to reality in our model
    in as far as the distance between the nuclei is of the order
    of $10^{-12}m$ in our model, so that a chain of atoms across the pearl 
    has a few thousand atoms in it.}
  \label{imbedded}
  \end{figure}



  \subsection{Our Picture of Dark Matter Pearls:}
\begin{itemize}
\item{\color{blue} Principle } Nothing but Standard Model!
  {\em (seriously would mean not in a BSM-workshop.)} 
\item{\color{blue} New Assumption} Several Phases of Vacuum with Same Energy
  Density \cite{MPP1, MPP2, MPP3, MPP4, tophiggs, Corfu1995}.
\item{\color{blue} Central Part} Bubble of New Phase of Vacuum with E.g. carbon
  under very High Pressure, surrounded by a surface with tension $S$
  (=domain wall) providing the pessure.
\item{\color{blue} Outer part} Cloud of Electrons much like an ordinary Atom
  with a nucleus having a charge of order
  of the number of electrons in the thin layer of electrons outside
  the genuine bubble region.
\item{{\color{blue}Even further out }} Further out one finds the
  dust collected by the pearl
  (through milliards of years) with the hardening and densification due to
  the effect of the unusually high homolumo gap.
\end{itemize}
The structure of our dark matter particles is illustrated in Figures 
\ref{dimensions} and \ref{imbedded}.

      \section{Achievements}

\subsection{Fitting the Velocity Dependence of the Ratio $\frac{\sigma}{M}$}

One somewhat supporting property of our model is that we can get it to match
crudely with the actual interaction of dark matter particles as seemingly
needed to fit the star velocities especially in dwarf galaxies, as estimated by
Correa \cite{CAC}. 
  
\begin{figure}
  \includegraphics[scale=0.9]{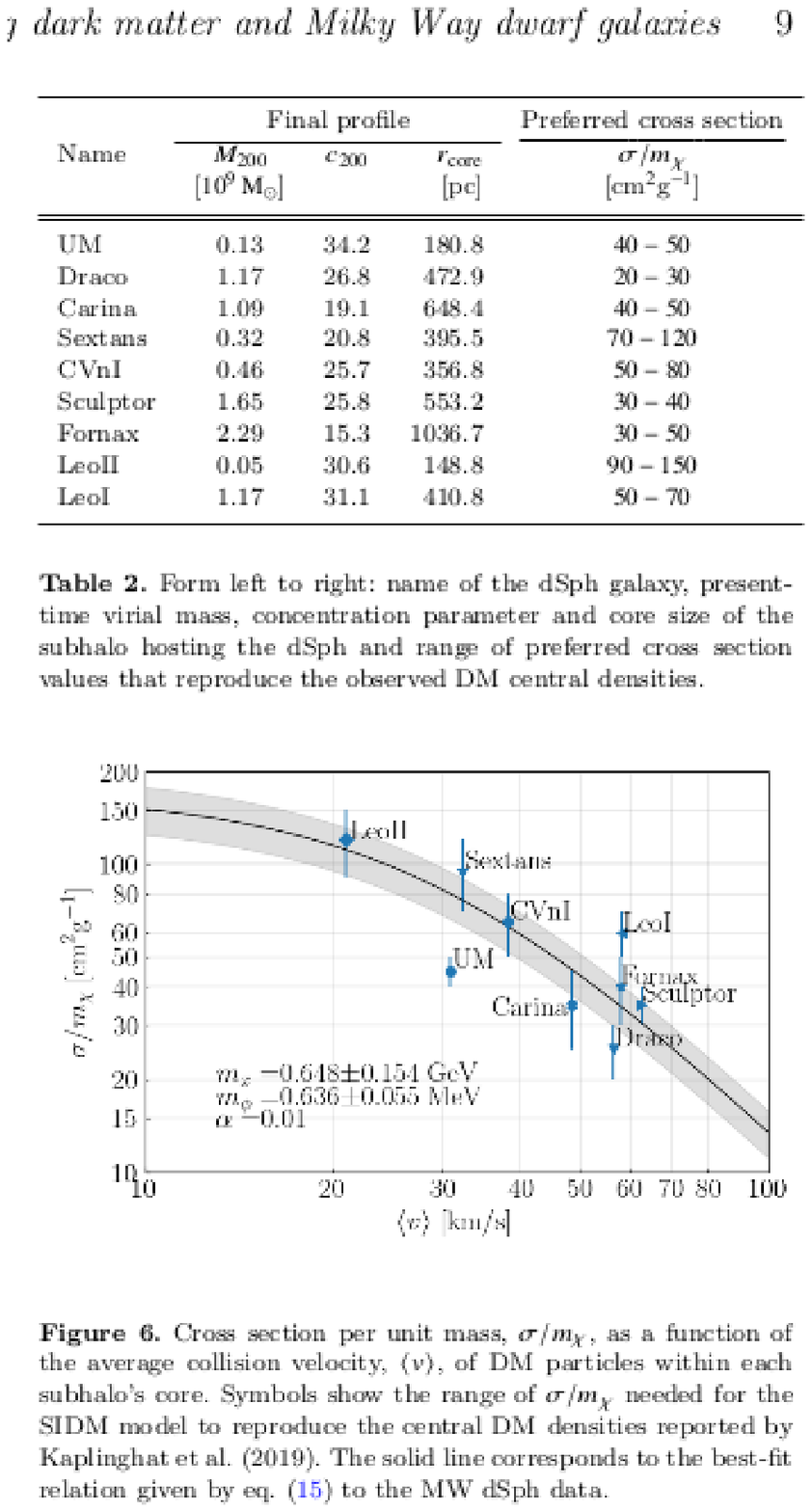} 
  \caption{Extract from Correa's paper illustrating the fits
    of the cross section to mass ratio obtained for different dwarf
    galaxies, as function of the estimated velocity for the relevant galaxy.}
   \label{Correa}
  \end{figure}
\begin{figure} 
  \includegraphics[scale=0.65]{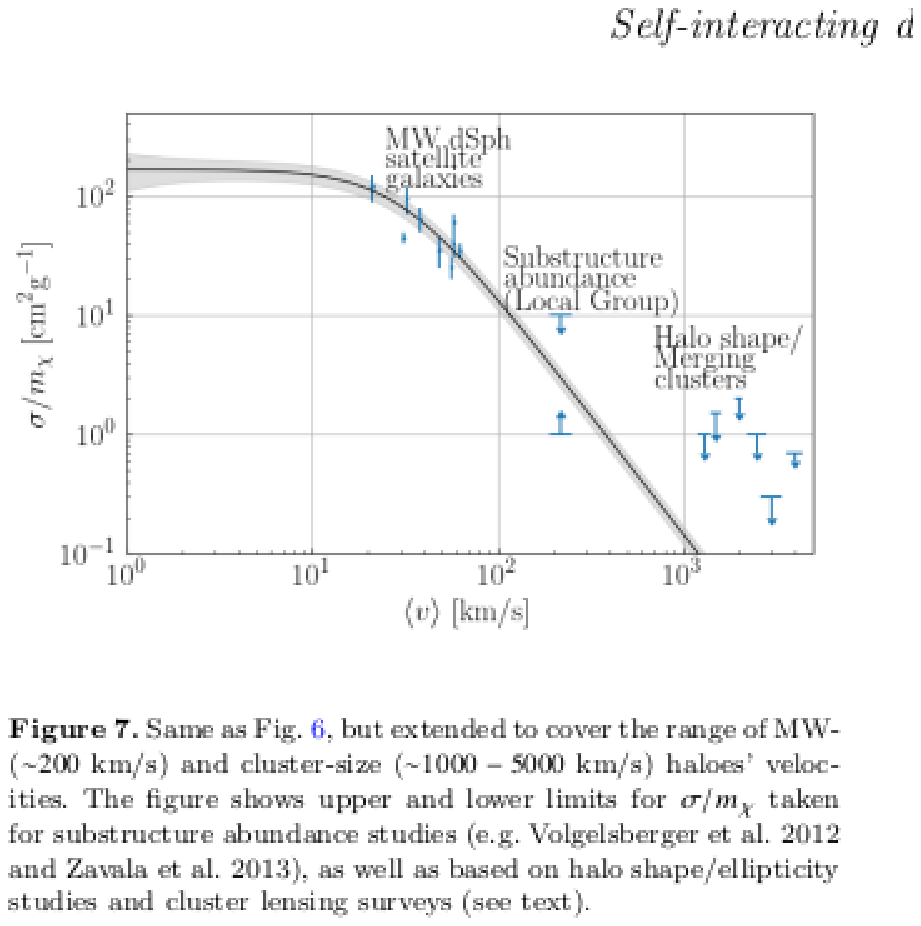}
  \caption{Same as Figure \ref{Correa} but extended to cover the range of Milky
  	Way and cluster-size haloes' velocities.}
  \label{extended}
  \end{figure}

  The a priori story, that dark matter has only
  gravitational interactions, seems not to work perfectly \cite{annika, SIDM, firstSIDM}: Especially in dwarf
  galaxies (around our Milky Way), where dark matter moves slowly, an appreciable
  cross section to mass ratio $\frac{\sigma}{M}$ seems to be needed
  going in the limit of low velocity to $150cm^2/g$, according to the fits in 
  Figures \ref{Correa} and \ref{extended}.
  
  We believe that our bubbles or pearls have become dirty in the sense that
  they are typically surrounded by a dust grain much like the dust grains
  that exist in outer space. Then, if as we shall
  guess (and get confirmed) the dust grain is appreciably larger than
  the bubble, it will be the size of this dust grain that determines the
  cross section $\sigma$ for the dark matter particle.
  The dark matter particles do not just end up being ordinary dust grains
  because according to big bang nucleosynthesis 
  there is not enough ordinary matter to make up all the dark matter.
  Only if the ordinary matter is packed so compactly inside pearls so as to 
  remain hidden already in the big bang nucleosynthesis
  era can we maintain a model in which the dark matter is basically
  ordinary matter. But we need to have the significantly bigger part
  ``packed away'' so that it is not seen as ordinary matter. So we must
  have the majority of the mass $M$ of our dark matter dirty particle
  sit inside the new vacuum bubble under high pressure, and not be free to show
  up as ordinary matter. So the full mass $M$ of the dark matter particles in  
  our model must be dominated by the mass of the little bubble, which thus
  essentially also has the mass $M$.

  Now one shall have in mind that since dark matter is only safely observed
  via its gravitational force on stars and galaxies and in the case of
  gravitational
  lensing on light, and since one looks over very large distances, it is
  impossible to distinguish by these main observations of dark matter whether
  there are many small dark matter particles or fewer but heavier ones. In fact
  it is thus only the cross section over the mass $\frac{\sigma}{M}$ which
  is observable.

  Correa \cite{CAC} fitted this ratio for a series of dwarf galaxies 
  and there were
  also other estimates of this ratio from galaxies and galaxy clusters
  giving some estimates for higher velocities.

  \subsubsection{Using the Dust grain to get a Mass}\label{dgm}
  If we took it as a crude estimate that the dust grain around our bubbles
  would have essentially the usual size $0.1 \mu m - 0.2 \mu m$ as for the dust without any
  dark matter inside, then we could use the low velocity value of the ratio
  \begin{eqnarray}
    \frac{\sigma}{M}|_{v\rightarrow 0} &=& 150 cm^2/g = 15 m^2/kg
    \end{eqnarray}
  to estimate the mass and the cross section separately
  \begin{eqnarray}
    \sigma &=& (0.2 \mu m)^2 = 4*10^{-14}m^2\\
    M &=& \frac{\sigma}{\frac{\sigma}{M}}\\
    &=& \frac{4*10^{-14}m^2}{15 m^2/kg} = 3*10^{-15}kg\label{dgmfM}\\
    &=& 1.6 *10^{12} \, GeV.
    \end{eqnarray}

  \subsubsection{The Size of the Bubble with New Vacuum}
  In section \ref{Xray} we shall mention, that we already \cite{theline,Corfu2019, Corfu2017,
  Bled20, Bled21} some time ago
  have fitted both the value 3.5 keV for the photon energy of an
  X-ray emission line found by satellite X-ray observations and
  not easily attached to any known line, and the intensity with
  which it was observed, by a single parameter combination
  $\frac{\xi_{fS}^{1/4}}{\Delta V} = 0.6 \, MeV^{-1}$. 
  In our model this quantity
  is rather trivially connected with the fermi momentum $p_f$
  of the degenerate electrons in the new vacuum, when filled with some ordinary
  matter so as to balance the pressure from the skin around the bubble of new
  vacuum. In fact this fermi momentum becomes with our fitted
  parameter combination
  \begin{eqnarray}
    p_f &=& 2/\frac{\xi_{fS}^{1/4}}{\Delta V} = \frac{2}
         {0.6 \, MeV^{-1}} =3.3 \, MeV.
    \end{eqnarray}

  Such a relativistic fermi momentum $p_f = 3.3 \,  MeV$ leads to a number 
  density of electrons and thus mass density:
  \begin{eqnarray}
    \rho_{ number \; e}&=& \frac{2*4\pi }{3*(2\pi)^3}*p_f^3
    \\
    &=& \frac{1}{3\pi^2}*p_f^3,
\end{eqnarray}
    which for two nucleons per electron leads to a mass density 
   \begin{eqnarray} 
    \rho_B&=& {2m_N}*\frac{1}{3\pi^2}*p_f^3\\
   &=&5.2 *10^{11}kg/m^3.
  \end{eqnarray}

  If the bubble shall have a mass $M = 3*10^{-15}kg = 1.6 *10^{12} \, GeV$ it shall have a radius $R$ given by
  \begin{eqnarray}
    \frac{4\pi}{3}R^3& =& M /\rho_B
    = 6*10^{-27}m^3\\
    \hbox{or } \qquad R &=& 1.2 *10^{-9}m = 1.2 \, nm\label{dgmfR}
  \end{eqnarray}

  Indeed with the dust grain radius being $0.2 \mu m$, the bubble inside would be
  about 200 times smaller in extension than the dust part. But if we speculate
  that the
  dust grain around the bubble is more hard to break into pieces than
  the normal dust grain, it could have grown somewhat bigger and then
  the dust grain size relative to the bubble would be even bigger than
  the here first mentioned factor 200.

  If we e.g. took it that the grain around the bubble would be say
  3.5 kev/ 3ev times bigger than the normal dust grain, then
  we would get a factor 200000 rather than only the 200 for the size of the
  dust grain in the dark matter pearl relative to its bubble of new vacuum.

    \subsubsection{The Effect of the Homolumo gap on the Dust Region}
\label{ehg}
Now we want to estimate the effect of the homolumo gap - as already
alluded to above in section \ref{IHGDM} -  on the
  dust region into which the bubble has come to be a seed.

  For this purpose we would like to get an idea about how the
  part of the wave functions for the filled states in the dust region looks.
  Now in this region in which nuclei for the dust are present we shall use,
  what one could call a tight binding approximation, in which we build up a wave
  function ansatz being a superposition of single atom eigenfunctions such as
  1s, 2s, 2p, ...That is to say we take ansatzes of the form
  \begin{eqnarray}
    \psi_{ans}(\vec{x})&=& \Sigma_{nucl.}\psi(\vec{x}(nucl.))*
    \psi_n(\vec{x}-\vec{x}(nucl.)), 
    \end{eqnarray}
  where $\psi_n$ is the eigenwave function for a certain level $n$
  in the atom corresponding to the nucleus type making up the dust,
  and the sum runs over the nuclei in the dust region, which are
  enumerated by the name $nucl.$ . The adjustable part of the
  ansatz is the smoothly varying (weight) function $\psi$.

  In the approximation that this $\psi$ is smooth the eigenvalue
  equation for such an ansatz $\psi_{ans}$ will be essentially the usual
  Schr{\o}dinger equation, just with an extra constant potential
  $V$ equal to the energy eigenvalue of the level $n$. If the energy
  eigenvalue for the level $n$ is $E_n$,
  then the effective Schr{\o}dinger equation
  formulated for the adjustable function $\psi$ in the ansatz
  becomes in the dust region
  \begin{eqnarray}
    -\frac{1}{2m_{eff}}\Delta \psi -|E_n|\psi &=& E\psi.  
    \end{eqnarray}
  Now because of the homolumo gap we expect that the eigenvalue
  for the filled states shall be negative and of the order -3.5 keV.
  Since this value -3.5 keV is a large number numerically compared to many
  of the $|E_n|$ values, the solution in the dust region for
  such a filled wave function will fall off exponentially as one goes further
  and further away from the central bubble.

  This then means that the wave functions for filled states only survive
  far out in the dust region, when $|E_n|$ is of the order of magnitude
  3.5 keV. That is to say only the levels with very strong binding will be
  filled for the atoms in the dust region.

\begin{figure}
  \includegraphics{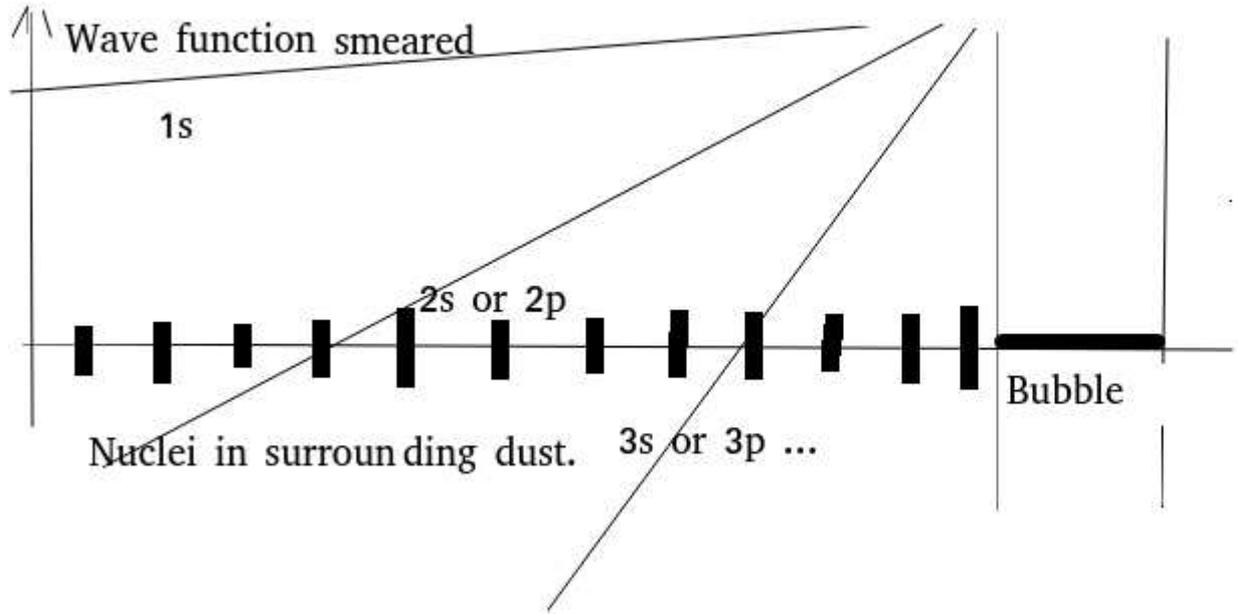}
  \caption{ \label{wavetpf} Here we imagine to make a wave function
    ansatz as a superposition of certain wave functions around the nuclei in the
    dust - symbolized by the short vertical thick lines in the left-hand region -
    but smeared out in the high nuclear density approximation. As the (local)
    wave function around the nuclei one can choose the 1s tightest bound state
    or the 2s or 2p states in the next level or 3s ....
    Since the eigenvalue for the electron energy is of the order of
    -3.5 keV (a numerically large number from the usual atomic physics point
    of view) the smoothed wave function describing the relative amplitudes for
    various atoms in the dust range will fall drastically as we go away from
    the bubble with its special and strong effects for the majority of the atom
    levels. Only if the binding to the
    state used for the ansatz is of order (-)3.5 keV will the smeared
    weighting wave be more
    flat. We imagine that this plot is logarithmic, so that the
    exponential fall off (towards the left in this figure) is represented
  by straight lines.}
\end{figure}

  The other levels in the dust atoms will only have a chance to be filled
  very close to the bubble - see Figure \ref{wavetpf}.
  
Having argued that only the lowest levels in the dust atoms are filled,
we expect that the electron clouds around the dust region nuclei become
much smaller than the usual size of such atoms. If the distance between the
atoms in the dust region is determined by the size of these clouds, then
the dust material would be expected to contract due to the here
studied effect of the homolumo gap for this material. Taking this
scaling to make the dust atoms smaller seriously, then we estimate
the scaling ratio to be of the order of the energy ratio of the
3.5 keV relative to a typical binding energy in the usual valence electron
level, say a few electron volts, say for simplicity 3.5 eV. Since the
Coulomb potential around a nucleus in the dust is proportional
to the inverse radius
\begin{eqnarray}
  V &\propto & 1/r,
  \end{eqnarray}
we expect the radius of the electron cloud around the nuclei in the
dust to be $\frac{3.5 keV}{3.5 eV} =1000$ times smaller in size.

This should then mean, that the density should be 
increased by a factor
1000 in each of the three dimensions.
We must, however, unfortunately admit that so strong a contraction of the
dust material is absurd in as far as it would make the hardened dust more
compact than the very bubble itself, which is supposed to cause this contraction.
The material in the bubble has namely, according to our fit to the 3.5 keV X-ray
line and its intensity, a density $5.2 *10^{11}kg//m^3$, which is only
about $10^8 = 464^3$ times bigger than a ``normal'' density, say
$5*10^3kg/m^3$. So a contraction of more than a factor of 464 in each direction
is at least unreasonable, and may even be unbelievably too much.

From similar dimensional arguments the energy of moving an atom closer
to a neighbour by a distance of the order of the distance to this neighbour
should go up by a factor 1000. Thus the force  arising from compressing
a cubic meter of the material by a strain of order unity would be
$1000^2=10^6$ times strengthened by the homolumo gap effect. But if we take
the little more reasonable value of at most a factor $464$ for each
direction, we would rather only get a $464^4 = 5*10^{10}$ times 
bigger elastic modulus for the hardened material than it had before.

We could by such dimensional arguments, using a unit system
in which the fine structure constant is conceived of as a velocity and
put to unity also taking the Planck constant $\hbar =1$, claim that
the length has dimension $[1/energy]$ and e.g calculate the
dimensionality of the elastic modulus $E$ of a material to be $[energy^4]$.
  Then using the dimensional argument and screwing up the energy
  by a factor $1000$, the modulus would go up by $1000^4 = 10^{12}$.
  But this is too much and we should not believe more than say
  $400^4 = 2.5*10^{10}$.

  Above we assumed that the modification of the dust material was so
  as to let the homolumo gap of 3.5 keV get extended from the bubble throughout 
  the dust grain. But that is presumably not true, 
  because the tails of the wave functions 
  reaching out in the dust region get thinner and thinner, so that the
  connection to the interior of the bubble gets very weak. Indeed in Figure 
  \ref{wavetpf} we sketch
  how the wave functions fall off exponentially with the distance
  to the bubble. As soon as the wave function truly feeling the bubble, meaning
  its homolumo gap effect, drops down exponentially and becomes small
  presumably its effect in modifying the atomic structures also gets small.
  So when even the wave functions dominating furthest out become small,
  we must expect that the appearance of relatively undisturbed standard atomic wave
  functions for atoms in the dust will be allowed by the Pauli principle. Thus
  the further out the dust atoms, the more and more normal they become and the
  effective homolumo gap is expected to diminish gradually with the distance from
  the bubble. In other words only the most nearby part of the dust 
  gets fully hardened.
  So to expect the correction factor $1000^4$ for the modulus $E$ should be
  considered rather an upper limit for the correction.

  Even the more moderate number $400^4=2.5*10^{10}$ is only an upper limit.

  Very far out we may expect quite normal dust, and we might take this as an
  argument to allow us to take the quite normal size $0.1 \mu m$ even for
  the dust grain around the dark matter bubble.

  But we should have in mind that if the hardening has an influence
  on the size, it is presumably by making the hardened dust grain less easy
  to break into pieces. Thus the hardened grain will survive without breaking up
  more easily than the normal dust grain. That of course will tend to make the
  dust grain around the dark matter bubble larger than the normal dust grain.

  \subsubsection{$\frac{\sigma}{M}$ Velocity Dependence, Characteristic Velocity}
\label{vdp}
  When the collision velocity $v$ is very low of course the geometrical
  size of the dust grain around the bubble determines the cross section,
  because there is not enough energy in the collision to cause any appreciable
  penetration of the dust grains into each other.

  However, when the velocity gets higher the grains get deformed, but the
  bubbles are extremely heavy and in first approximations they are very hard to
  turn into a motion in another direction. Now it is exactly turning
  them into an appreciably different direction which is needed, in order
  that the collision will be significant for the motion of the dark matter
  pearls and give a cross section that can be detected in say the
  dwarf galaxy simulations and fits.

  Let us imagine the collision at higher velocity like this:

  \begin{itemize}
  \item In first approximation - of high velocity - the bubbles
    which come in with a relative impact parameter $b$ just move undisturbed
    through with undisturbed velocity until they have
    compressed the dust grains so much that the energy of the deformation
    has become of the order of the incoming kinetic energy of the bubbles.
    (Since the bubbles are in mass supposed to dominate, essentially all the
    kinetic energy sits in the bubbles.)
    \item We describe the deformation of the dust grains as the displacement
    of the bubble inside the grain as far as is needed, assuming the
    elastic deformation of the grain going on mainly around the bubble
    requires the least amount of elastic energy.

  \item The dust grains themselves are in our crudest approximation
    considered so light compared to the bubbles, that they can be accelerated
    and decelerated without any significant energy use. So the grains must
    be imagined to pass through each other when the impact parameter $b$ is
    smaller than the sum of their radii by the dust particles  
    going a bit aside so as to be able
    to pass. Then they immediately turn back to essentially the original orbits
    because 
    the small but heavy bubbles force
    them to return to their original orbits, unless of cause the bubbles have
    themselves been stopped and brought out of their straight trajectories during 
    the detour.

  \item The estimation of the elastic energy stored during the passage
    of the two dark matter particles is thus mainly due to - the square, one
    could say, of -  the displacement during the passage of the bubbles
    relative
    to their associated grains of dust.

    At the most stressed point during this passage the displacement distance is
    \begin{eqnarray}
      \hbox{`` displacement''}&=& r_{dust}-b/2,
    \end{eqnarray}
    where $r_{dust}$ is the radius of the dust grains. In reality the
    dust grains are not at all spherical, so that the radius is only a very
    crude concept and also they will typically not all be of the same
    radius even to the approximation that they are spherical.

  \item The crudest way to estimate the elastic energy from this
    $r_{dust}-b/2$ displacement is to say that a volume of dust grain
    material at least pressed through is
    \begin{eqnarray}
      \hbox{ `` volume'' } &=& \pi *R^2*(r_{dust}-b/2),
    \end{eqnarray}
    where $\pi *R^2$ is the area for the heavy bubble.
    
    The corresponding elastic energy may crudely be estimated to be 
    \begin{eqnarray}
      \hbox{ ```elastic energy ''}&=& E *R^2*(r_{dust}-b/2)^2/b.
    \end{eqnarray}
   The $b$ in the denominator is the ``original'' or rather ``final''  length of the
    column being compressed by the amount $r_{dust}-b/2$
    (if you wish you can consider it that the length of the
    column of dust matter compressed has the length $b/2$ and that this
    factor of a half is cancelled by the 1/2 in the potential energy; 
    or just take it as an order of magnitude estimate). 
    
    \end{itemize}

  Really this expression for the elastic energy does not quite correspond to
  a correct picture of the deformation of the dust grain, in order to allow
  the two dust grains to pass each other without any deviations from the
  unchanged velocity motion of the very heavy bubbles. 
  This means of course that the bubbles must be displaced relative to
  its dust grain so as to come to only a distance $b/2$ from the
  surface of the grain. This not so correct deformation pattern is thought
  to be that of the dust material in the very neighbourhood of the column of
  dust
  with length of the order $b/2$ - the most compressed part being of this
  length - and cross section $\pi * R^2$, the cross section of the bubble.
  But it is very likely that one can find a more smoothed out deformation 
  with a somewhat lower energy. 
  The true deformation should be the one with the minimal
  elastic energy, but still with the correct displacement of 
  the bubble relative to the 
  dust grain. Thus really the above order of magnitude estimate for the
  elastic energy is rather an upper bound. So rather we have
   \begin{eqnarray}
      \hbox{ ```elastic energy ''}&\le& E *R^2*(r_{dust}-b/2)^2/b.
    \end{eqnarray}
   But we may still believe that order of magnitudewise it is rather close
   to equality.

   The condition on the kinetic energy $\frac{1}{2}M*v^2$ of a dark matter particle with mass dominated by that of the bubble $M$ 
   and velocity $v$, which is just
   barely being stopped in such a collision, is provided order of 
   magnitudewise by
   \begin{eqnarray}
     \frac{1}{2} M *v^2 &\approx & E*R^2*(r_{dust}-b/2)^2/b\\
     \hbox{or } \qquad \sqrt{\frac{M}{2E*R^2}}*v &\approx & \frac{r_{dust}-b/2}{\sqrt{b}}\\
     &=& r_{dust}/\sqrt{b} - \sqrt{b}/2.\\
     \hbox{or } \quad (\sqrt{b})^2 + \sqrt{\frac{2M}{E*R^2}}*v*\sqrt{b} -2r_{dust}&\approx& 0
   \end{eqnarray}
   It should be borne in mind that these equations are only 
   relevant for the impact parameter $b\le 2r_{dust}$, since for larger
   $b$ of course the dust grains do not even touch. For the variable
   in the second order equation, the square root of the impact parameter
   $\sqrt{b}$, this relation for the interesting
   cases mean $\sqrt{b}\le \sqrt{2r_{dust}}$. The solution to this second
   order equation is
   \begin{eqnarray}
     \sqrt{b} &=& -\sqrt{\frac{M}{2E*R^2}}*v
     \pm \sqrt{\frac{M}{2E*R^2}*v^2+2r_{dust}}
   \end{eqnarray}
   We can check that for very low velocity $v\approx 0$ we of course obtain the
   cross section corresponding to the impact parameter of just stopping
   \begin{eqnarray}
     \sigma = \pi b^2 &=& \pi(\sqrt{2r_{dust}})^4\\
     &=& \pi *4 r_{dust}^2.
   \end{eqnarray}
   For the purpose of estimating the ratio $\frac{\sigma}{M}$ it is more
   suitable to write the equation for the combination 
   $\frac{\sqrt{b}}{M^{1/4}}$, i.e.
   \begin{eqnarray}
     \frac{\sqrt{b}}{M^{1/4}} &=& - \sqrt{\frac{\sqrt{M}}{2E*R^2}}*v\pm 
     \sqrt{\frac{\sqrt{M}}{2E*R^2}*v^2+\frac{2r_{dust}}{\sqrt{M}}}.
   \end{eqnarray}


   In the case of low velocity we can make the approximation of dropping
   the  $v^2$ term under the square root and obtain
    \begin{eqnarray}
     \frac{\sqrt{b}}{M^{1/4}} &=& - \sqrt{\frac{\sqrt{M}}{2E*R^2}}*v\pm 
     \sqrt{\frac{2r_{dust}}{\sqrt{M}}} \qquad\hbox{ ( for small $v$)}\\
     &=& \sqrt{\frac{2r_{dust}}{\sqrt{M}}}*\left (1-v*\sqrt{\frac{M}
       {2r_{dust}2E*R^2}}\right )\quad \hbox{(using of course $+$ for $\pm$ ) }\\
     &=& \sqrt[4]{\frac{\sigma}{M\pi}|_{v\rightarrow 0}}\left (1- \frac{v}{v_0}
     \right ),     
    \end{eqnarray}
    where
    \begin{eqnarray}
      v_0 &=& \sqrt{\frac{4r_{dust}E*R^2}{M}}.
      \label{v0}
      \end{eqnarray}
    Taking the fourth power we then obtain
    \begin{eqnarray}
      \frac{\sigma}{M}&=& \pi \left ( \frac{\sqrt{b}}{M^{1/4}} \right)^4\\
      &=& \frac{\sigma}{M}|_{v\rightarrow 0}*(1-v/v_0)^4 \label{formula}.
    \end{eqnarray}

    Using the data of Figure \ref{Correa} from Correa \cite{CAC} and 
    fitting to the formula (\ref{formula}) above for $v=100$ km/s we obtain
    $v_0=230$ km/s. 
    Taking the smaller velocity $v=35$ km/s to fit we get 220km/s.
    
    The value of the fermi momentum $p_f =3.3$ MeV determines the mass
    density of the material in the
    bubble of the second type of vacuum:
\begin{eqnarray}
    \rho_B &=& \frac{2m_N}{3\pi^2}p_f^3 = 5.2*10^{11}kg/m^3.
    \label{rhoB}
\end{eqnarray}
So  we have
\begin{eqnarray}
  \frac{4\pi}{3}*R^3*5.2*10^{11}kg/m^3  &=& M.
\end{eqnarray}
and therefore
\begin{eqnarray}
 \frac{R^2}{M} &=& \frac{1}{\sqrt[3]{M}} *
  (\frac{3}{4\pi*5.2*10^{11}kg/m^3})^{2/3}\\
    &=& \frac{1}{\sqrt[3]{M}*1.68*10^8kg^{2/3}/m^2}
\end{eqnarray}

Taking $v_0=220km/s = 2.2*10^5m/s$ we thus obtain from equation (\ref{v0}):
\begin{eqnarray}
  2.2*10^5m/s &=& \sqrt{\frac{4r_{dust}E}{\sqrt[3]{M}*1.68*10^8kg^{2/3}/m^2}}\\
  \hbox{ or } \qquad 4r_{dust}E &=& (2.2*10^5m/s)^2*1.68*10^8kg^{2/3}/m^2*\sqrt[3]{M}\\
  &=& 8.1*10^{18}kg^{2/3}/s^2*\sqrt[3]{M}.
\end{eqnarray}

 For say $M=10^{-15}kg$, we then obtain:  
\begin{eqnarray} 
 4r_{dust}E &=& 8.1*10^{13}kg/s^2.
\end{eqnarray}

Without a story about the influence of the homolumo gap effect this
number for $4r_{dust}E$ is far too high. In fact without the
homolumo gap modification we expect
\begin{eqnarray}
  E &\approx & 10^9 N/m^2 \qquad \hbox{ (usual materials)}\\
  r_{dust}&\approx & 10^{-7}m \qquad \hbox{(typical cosmic dust grain)}\\
  \hbox{ giving } \qquad  4Er_{dust} &\approx & 400 N/m.
  \label{Er} 
\end{eqnarray}
(Have in mind $N/m =kg/s^2$).

However in section \ref{ehg} we suggested a possible homolumo gap
correction factor to the elastic modulus $E$ of $1000^4= 10^{12}$ or perhaps a
little more reasonable factor of $400^4=2.5*10^{10}$. 
Assuming $r_{dust}$ is not changed these two possible homolumo gap 
corrections increase $4Er_{dust}$ up to
\begin{eqnarray}
  4Er_{dust}|_{\hbox{homolumo corrected}}&=& 400N/m*10^{12}
  = 4*10^{14}N/m
\end{eqnarray}
and
\begin{eqnarray}
	 4Er_{dust}|_{\hbox{homolumo corrected}}&=&
  400N/m*2.5*10^{10}
  = 1*10^{13}N/m 
\end{eqnarray}
respectively.
These values are to be compared with the value $4Er_{dust}= 8.1*10^{13}$ N/m
obtained in equation (\ref{Er}) using a characteristic velocity 
$v_0=220$ km/s and pearl mass $M=10^{-15}$ kg.
 
Clearly we really need such a homolumo gap correction to even get just 
crude agreement.

Formally we could adjust the mass $M$ to make the estimates agree completely
and declare we had fitted the mass to be the third power of the deviation ratio
5 bigger than the $10^{-15}kg$, thus getting $M \approx 10^{-13}kg$, but of
course it makes little sense, since the dependence on the mass $M$ is so
weak for this fitting of the velocity dependence, coming essentially
via its sixth root. The uncertainty gets huge.

\subsection{Mass Discussion}
\label{ndg}
In the philosophy that the dark matter particles are provided with
dust grains around them, it is of course clear that the density in
the universe of such dark matter particles should better
be smaller than the number of dust grains estimated from observations,
since otherwise our model would predict more dust grains than
observed.

If we at first ignored the suspicious corrections of the density of the
dust material and just assumed that each dark matter bubble were associated
with a dust grain of usual density $\sim 2000 kg/m^3$ and size
$r_{dust}=0.1 \mu m = 10^{-7} m$, we could estimate
the grain number density - very crudely -
by taking it that 10-logarithm of the ratio DGR  of dust mass to
gas mass is \cite{DGR}
\begin{eqnarray} 
  log DGR &=& 2.45 log\frac{Z}{Z_{\odot}} -2.03.
\end{eqnarray}
Here $Z$ is the metallicity in the region in question (i.e. the amount
of heavier than helium elements relative to all elements) while
$Z_{\odot}$ is this metallicity in the surface of the sun ($\odot$).

Let us crudely estimate:
\begin{eqnarray}
  \hbox{`` mass of typical grain''}&=& 2000kg/m^3 *(10^{-7}m)^3\\
  &=& 2*10^{-18}kg\\
  \hbox{`` critical density ''}= \frac{3H^2}{8\pi G}&=& 10^{-26}kg/m^3\\
  \hbox{ `` ordinary matter density ''} &=& 5\% \hbox{ of } 10^{-26}kg/m^3\\
  &=& 5*10^{-28}kg/m^3\\
  \hbox{ hereof say ``gas'' }&=& 3*10^{-28}kg/m ^3.\\
  \hbox{ If $Z=Z_{\odot}$ then } DGR &=& 0.009,\\
  \hbox{and ``dust density''}\rho_{dust} &=& 3*10^{-28}kg/m^3*0.009\\
  &=& 2.7*10^{-30}kg/m^3\\
  \hbox{``Typical grain number density''}&=& \frac{2.7*10^{-30}kg/m^3}
       {2*10^{-18}kg}\\
       &=& 1.4 *10^{-12}m^{-3}.
\end{eqnarray}
This would correspond to a typical distance between the grains of 11 km.

The dark matter average density is similarly
\begin{eqnarray}
  \hbox{``DM density''}&=& 24 \% \hbox{ of } 10^{-26}kg/m^3\\
  &=& 2.4*10^{-27}kg/m^3.
\end{eqnarray}
So with a dark matter pearl mass $M$ the number density of the dark
matter pearls is
\begin{eqnarray}
  \rho_{\hbox{DM number}}&=& \frac{2.4*10^{-27}kg/m^3}{M}.
\end{eqnarray}

If this number density should be equal to that of the dust grains, we
should have
\begin{eqnarray}
  \frac{2.4*10^{-27}kg/m^3}{M}&=& 1.4 *10^{-12}m^{-3}\\
  \hbox{leading to } \qquad M &=& 2*10^{-15}kg.
  \end{eqnarray}

Of course there would be a lot of dust grains without any dark matter bubble
in it, and thus we should consider the number $2*10^{-15}kg$ as only a
{\em lower} limit for the mass $M$ of the dark matter particle:
\begin{eqnarray}
  M&\ge& 2*10^{-15} kg= 1.2*10^{12}GeV\label{ndgfM}.
  \end{eqnarray}
Corresponding to such a mass we have a radius
\begin{eqnarray}
  R &\ge & \sqrt[3]{\frac{2*10^{-15}kg}{\frac{4\pi}{3}*5.2*10^{11}kg/m^3}}\\
    &=& 1.0*10^{-9}m\label{ndgfR}
\end{eqnarray}

Below in  section \ref{fty} we shall present yet another lower bound 
for the mass $M$ of the dark matter pearls: $3.9*10^{13} GeV$, see
equation (\ref{ftyfMc}).

{\em That order of magnitudewise, including the speculation of the
  hardened dust material, we can achieve a consistent dark matter picture as
  these dirtified bubbles should be considered an achievement of our model.}

\subsection{ Resolving Seeming Disagreement of
    DAMA and
    Xenon-experiments.}
  

  Most underground experiments are designed to look
  for dark  matter particles hitting the nuclei in the experimental apparatus,
  which is then scintillating so that such nuclear recoil events can be seen.
  But our pearls are excited in such a way that they send out energetic
  {\em electrons} (rather than nuclei), and thus does not match with what is
  looked for, except in the DAMA-LIBRA experiment, in which the only signal for
  events coming from dark matter is a seasonal variation due to the Earth
  running towards or away from the dark matter flow. Recently the Xenon1T
  experiment has actually observed an excess of electron recoil events, 
  which we suggest are coming from the decay of our excited dark matter pearls. 

  \subsubsection{Mass bound from need for Seasonal Oscillations in DAMA}
\label{fty}
  In order that our model shall be able to fully produce the DAMA-effect
  from the decay of our excited bubbles of new vacuum (or the like), it is
  needed that the bubbles or whatever, which produce the
  electron or gamma signal detectable by DAMA, come down to the experiment in less
  than a year. Otherwise of course the seasonal variation would be washed
  out and only a constant signal seen by the DAMA-LIBRA experiment.

  Now it is hard to estimate the terminal velocity for the bubbles
  penetrating solids, but we believe we can estimate the terminal velocity
  for the bubbles going through the liquid xenon in the Xenon1T experiment 
  due to the gravitational attraction of the Earth. 
  In fact there is a balance between
   the gravitational force $gM$ 
   and
  the Stoke's law force $6\pi \eta R v_{terminal\; Xe}$, so that in the fluid xenon
  the terminal velocity is given by
  \begin{eqnarray}
    Mg &=& 6\pi \eta R v_{terminal\; Xe}\label{terminal}.
  \end{eqnarray}
  In our earlier article \cite{extension} we took the viscosity of the liquid
  xenon to be
  \begin{eqnarray}
    \eta = \eta_{liquid \; gas}&\approx &100 \mu Pa\;s= 0.1mPa\;s,\label{eta}  
    \end{eqnarray}
  and obtained a limit
  \begin{eqnarray}\label{frext}
    M &\ge& 2.1 *10^{-15}kg = 1.2* 10^{12}GeV.
    \end{eqnarray}
  In fact  we here used the  experimental fact that there are 250 times more
  events seen in the DAMA experiment per kg scintillator than in the
  Xenon1T excess of electron recoil events. This we take to mean that
  the terminal velocity in the NaI of the DAMA experiment - and we extend it
  to all solids - is 250 times smaller than in the fluid xenon.
  
  Using equations (\ref{terminal}) and (\ref{eta}) together with
  \begin{eqnarray}
    M&=& \frac{4\pi}{3}\rho_B R^3,
  \end{eqnarray}
  and the density of the bubble $\rho_B = 5.2*10^{11}kg/m^3$ taken from equation (\ref{rhoB}), we obtain
  \begin{eqnarray}
    v_{terminal\; Xe} &=&\frac{gM}{6\pi \eta_{fluid \; gas}R}\\
    &=& \frac{2g \rho_BR^2}{
      9\eta_{fluid \; gas}}\\
    &=&
    1.13*10^{16}m^{-1}s^{-1}*R^2.
  \end{eqnarray}
  Taking the terminal velocity in solid material to be 250 times smaller
  \begin{eqnarray}
    v_{terminal \; solid}&\approx & \frac{v_{terminal \; Xe}}{250}\\
    \hbox{we get } \qquad v_{terminal\; solid}&\approx &
    4.5*10^{13}s^{-1}m^{-1}*R^2.
    \end{eqnarray}

  Taking the depth penetrated into the earth so as to reach the DAMA
  experiment to be 1400 m, the pearls need to come down in less than a year and
  thus to have a (terminal) velocity larger than $1400 m/1\, \hbox{year}$, i.e.
  \begin{eqnarray}
    v&\ge& \frac{1400m}{365*24*60*60s}\\
    &=& 4.4*10^{-5}m/s,
    \end{eqnarray}
  the requirement on the radius would be 
  \begin{eqnarray}
    R^2 &\ge &\frac{4.4*10^{-5}m/s}{
      4.5*10^{13}s^{-1}m^{-1}}\\
    &=&
    1.0*10^{-18}m^2\\
    \hbox{giving } \qquad R &\ge&
    1.0*10^{-9}m\label{ftyfR}
    \end{eqnarray}
   The corresponding condition on the mass $M = \frac{4\pi}{3}\rho_BR^3$ is
   \begin{eqnarray}
   	M &\ge& \frac{4\pi}{3}*5.2*10^{11}kg/m^3*(10^{-9}m)^3\\
   	\hbox{giving} \qquad M &\ge& 2.1*10^{-15}kg =1.2*10^{12} \, GeV. \label{ftyfM}
   \end{eqnarray}
  This agrees with the bound (\ref{frext}) from the foregoing article
  \cite{extension}.

It would be reasonable to correct this bound, obtained assuming that the
whole impact of dark matter is seasonally varied, to instead using the
more realistic assumption that only about $10\%$ of the incoming dark
matter is varying, since the ratio of the earth velocity 30 km/s to the
velocity of the solar system moving through the average dark matter  
300 km/s is about 10\% . The electron recoil excess in the Xenon1T experiment
of course contains both the varying and the constant part. So the factor 250, which
we used above and required the ratio of the terminal velocities of the dark matter in the two experiments to be, should rather have been 250/10\% = 2500. 
We require the speed in the solid $v_{terminal \; solid}$ to be unaltered
- it is given by the one year requirement. So we have to increase 
the $v_{terminal \; Xe}$ by
a factor 10 more. That would then increase the boundary value for the radius
of the bubble $R$ by a $\sqrt{10}=3.1$ and the boundary mass $M$ by a factor
$10^{3/2}=31$. So the corrected bounds become:
\begin{eqnarray}
	R&\ge&3.1*1.0 *10^{-9}m = 3.1*10^{-9}m\label{ftyfRc}\\
	\hbox{and } M&\ge& 31*2.1*10^{-15} kg =6.5*10^{-14}kg\label{ftyfMc}\\
	&=& 3.9*10^{13} \, GeV. 
\end{eqnarray}
  


\subsection{ The 3.5 keV X-ray Radiation}
\label{Xray}
  \begin{itemize}
  \item{\bf \color{blue} The Intensity of 3.5 keV X-rays from Clusters etc.}
    We fit the very photon-energy 3.5 keV and the overall intensity from a
    series of clusters, a galaxy, and the Milky Way  Center with one parameter
    $\frac{\xi_{fS}^{1/4}}{\Delta V} = 0.6 MeV^{-1}$ \cite{theline,Bled20}.
    \end{itemize}
    
   Even though we only need the one parameter
    $\frac{\xi_{fS}^{1/4}}{\Delta V}=\frac{2}{p_f}$, it is nice to know the
    notation:

  \begin{eqnarray}
    \Delta V &=& \hbox{`` difference in potential for a nucleon}\nonumber\\
    && \hbox{ between inside
      and outside the central part of the pearl''}\nonumber\\
    &\approx & 2.5 MeV\\
    \xi_{fS}&=& \frac{R}{R_{crit}}\hbox{estimated } \approx 5\\
    \hbox{where } R &=& \hbox{``actual radius of the new vacuum part''}
    \nonumber\\
    &\approx & r_{cloud \; 3.3 MeV}\\
    \hbox{ and } R_{crit}&=& \hbox{`` Radius when pressure so high}\nonumber\\
    && \hbox{ that nucleons
      are just about being spit out''}
    \end{eqnarray}
 
  \subsection{  About the Xenon1T and DAMA-rates:}
  \label{absrates}
  
We estimate the event rates seen by DAMA-LIBRA and in the Xenon1T electron 
recoil excess using energy considerations based on the kinetic energy of the 
incoming dark matter as known from astronomy. 
  In order to explain these 
  estimates it is necessary
  to know how we imagine
  the dark matter to interact and get slowed down in the 
  air or in the earth shielding the experiments,
  and how the dark matter particles get excited and emit 3.5 keV radiation
  or electrons.

    Our estimate of the absolute rates for the
    two experiments are very very crude because we assume that the dark matter
    particles - in our model small macroscopic systems with
    of the order of $10^{12}$
    of nuclei inside them - can have an exceedingly smooth distribution of
    lifetimes on a logarithmic scale.
    These calculations are discussed in \cite{extension}.
    Let us shortly review them here:

    The dark matter pearls come into the atmosphere with high speed 
    (galactic velocity)
    and are likely to be essentially stopped in the air, meaning they have slowed 
    down to a speed  $\sim$ 49 km/s below which collisions with nuclei can 
    no longer excite the 3.5 keV excitations. We consider the 
    alternative scenario where the stopping takes place in the Earth in 
    section \ref{earth}.
    
      Because the atmosphere roughly scales in density, rising by a
      factor $e$ for each $7km$ one goes down, the order of magnitude
      of the ``stopping length'' (meaning the distance over which the essential
      stopping of a pearl in our model takes place) cannot be much
      different from 7 km. Thus the corresponding ``stopping time", when the 
      incoming velocity is $\sim$ 300 km/s, is given  
      in order of magnitude by 0.023 s.

      We assume that there is a wide range of different 3.5 keV 
      excitations possible
      with different decay rates distributed smoothly in the logarithm
      of the decay time. It follows that
      the decay rates dominating for pearls that 
      survive for a time $t$ after their excitation before they are seen to decay
      will correspond to a life time of order $t$. Because of
      the tightness of the bound (\ref{ftyfM})
      for the mass $M$ from the requirement of
      the pearl reaching down faster than in a year and the estimated mass 
      (\ref{dgmfM}) from
      the dwarf galaxy self interaction estimate for low velocity, we must
      believe that the passage time $t$ is close order of magnitudewise
      to 1 year.

      Thus we have
      \begin{eqnarray}
        \hbox{``stopping time''} &\approx &\frac{7\; km}{300 \, km/s}\\
        &=& 0.023 s\\
        \hbox{``passage time''} &\approx & 1 \, year = 3*10^7 s
      \end{eqnarray}
      By essentially a dimensional argument, supported by assuming that
      excitations surviving longer before decaying also take more time to
      get excited, we get that there is a suppression factor $suppression$
      for getting
      the long living excitations excited and given by
      \begin{eqnarray}
        suppression &\approx & \frac{\hbox{``stopping time''}}
        {\hbox{``passage time''}}\\
        &\approx & \frac{0.023 s}{3*10^7 s}\\
        &=& 6*10^{-10}. \label{suppression}
        \end{eqnarray}
      Note that this theoretical $suppression$ ratio is independent of the pearl
      mass. It is to be compared to 
      the ratio of the 
      energy deposition rate in DAMA relative to the power from 
      the kinetic energy of the dark matter, making the crude 
      assumption that it is distributed uniformly over the matter down 
      to the 1400 m depth of the DAMA experiment. The energy deposition rate  
      per kg of scintillation material is calculated to be 
      $\hbox{``deposited rate"} = 2.7*10^{-22}W/kg$,
      assuming that all the DAMA-LIBRA events are
      due to decays with decay energy 3.5 keV. 
      Divided into a layer of 1400 m thickness
      the kinetic energy rate we found in \cite{extension} to be
      $\hbox{``power to deposit''} = 1.7*10^{-12} W/kg$. 
     We then obtain the experimental suppression rate 
      \begin{eqnarray}
      	suppression_{DAMA} &=&
        \frac{\hbox{deposited rate}}{\hbox{``power to deposit''}} \\
        &=&
            \frac{1.7*10^{-12}W/kg}{2.7*10^{-22}W/kg}\\
            & =& 1.6*10^{-10}.
      \end{eqnarray}
       This agrees surprisingly well with the crude theoretical 
       $suppression$ factor (\ref{suppression}) above.            
           
            Similarly for Xenon1T
        \begin{eqnarray}     
            suppression_{Xeon1T} &=& 6*10^{-13}.
        \end{eqnarray}

      Below in (\ref{st109}, \ref{st110}) we find the stopping times
      for the presumably less realistic case that the stopping first
      takes place in the stone. These stopping times are of the order of
      $10^{-6}s$ and thus shorter than the one based on the stopping in the air
      by about 4 orders of magnitude, i.e. they are $10^4$  times smaller,
      and thus will give $10^4$ times less rate prediction in DAMA.

  The ratio of the rates in the
    two experiments  - Xenon1T electron recoil excess  and DAMA - should in
    principle be very accurately predicted in our model, 
    because they are supposed
    to see exactly the same effect just in two different experiments
    in the same underground laboratory below the Gran Sasso mountain!
    One would therefore expect the rates to be the same, but the Xenon1T rate
    is about 250 times smaller than the DAMA rate. As mentioned in section \ref{fty}, we interpret this to mean that the terminal velocity in the 
    solid NaI of the DAMA experiment to be 250 
    or 2500 times smaller than in the fluid 
    xenon.

  \subsection{Jeltema and Profumo's Observation}
  \label{Jelt}

  Let us briefly mention an at first very strange observation \cite{Jeltema} from
  the point of view of the hypothesis that the 3.5 keV X-ray line observed 
  astronomically should indeed 
  come from dark matter: Jeltema and
  Profumo saw this 3.5 keV line in the X-ray spectrum from the Tycho
  supernova remnant! This is remarkable because otherwise one only sees this line
  coming from huge collections of stars such as galaxy clusters with supposedly a
  huge amount of dark matter, while the amount of dark matter in the supernova 
  remnant is expected to be relatively minute.

  In an earlier article \cite{Bled20} we assumed that our pearls could be excited by
  cosmic rays - of which there is an appreciable amount in supernova remnants -
  hitting the nuclei in the bubble of the new vacuum. If as we assumed the
  density and the size of this bubble is not so big that the nuclei
  significantly shadow for each other, the cross section for such a nucleus
  hitting collision divided by the mass will be just the value for this
  ratio obtained for a cosmic ray collision with a single nucleus. 
  Then we took it that the energy
  collected from the cosmic ray by the dark matter in the neighbourhood of the
  supernova remnant would mainly go into 3.5 keV excitations, of which in turn
  about a fraction $\alpha$ (the fine structure constant) would be
  emitted as X-rays. Such a model gave good agreement with the
  radiation observed by Jeltema and Profumo.
  
\section{Impact}

  \subsection{Illustration of Interacting and Excitable Dark Matter Pearls}

  The dark matter pearls come in with high speed (galactic velocity), but
  get slowed down to a much lower speed by interaction with the 
  atmosphere or the shielding
  mountains, whereby they also get excited to emit 3.5 keV X-rays or
  {\em electrons}. As mentioned above in section \ref{absrates}, we  
  believe that the essential stopping of the pearls is most likely to take place 
  in the atmosphere, but here we shall consider the possibility that it happens 
  in the earth above the underground experiments.
  
  Presumably already in the atmosphere the pearls with their dust around them will
  be heated so much that the dust grain will be burned off - much like
  the tail of a comet gets burned off by solar radiation. This means
  that a dark matter object penetrating into the Earth 
  becomes much smaller than the
  object out in space for which Correa estimated the cross section
  over mass ratio from the dwarf galaxy studies.
  
\begin{figure}
    \includegraphics{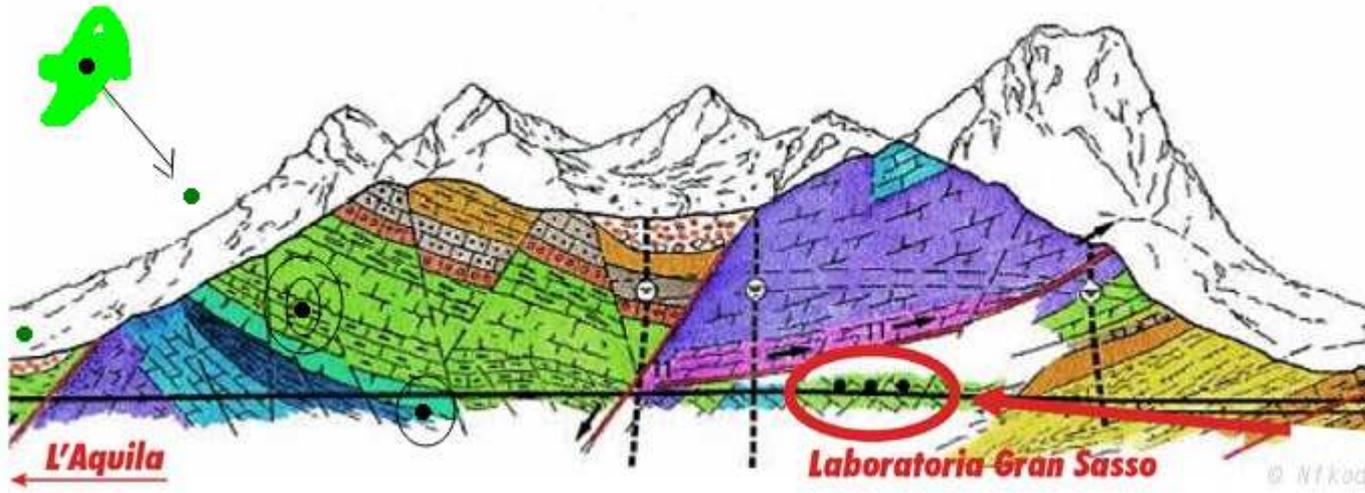}
    \caption{\label{ipn} On this figure the motion of a single
      dark matter particle as time goes on and it comes deeper down towards
      the Earth is illustrated as a series of pictures of the particle.
      Out in space from where it comes it still has its grain of dust
      around it. Then that gets stripped off, and at the same time the
      ``naked'' bubble gets excited, first strongly but as it goes on it
      de-excites most of its excitations - i.e. most of its internal energy -
      and deeper in only a very little part is left. In most cases
      actually the excitation energy is used up rather high up in the shielding
      or in the air. But somewhat seldomly an excitation survives down
      to excite the DAMA counter, or make an event to be counted as an
      Xenon1T electron recoil excess event.}
    \end{figure}

  The impact is illustrated in Figure \ref{ipn}. 

  \subsection{\bf Pearls Stopping and getting Excited in Earth Shield}
  \label{earth}

  What happens when the dark matter pearls, in our model of
  nano-meter size bubbles surrounded by dust grains, 
  hit the earth shielding above the experimental halls of e.g.
  DAMA?
  
  \begin{itemize}
  \item{\color{blue} Stopping} 
   %
    For very large velocities $v$ the dragging force acting on a pearl
    of radius $R$ (say a naked bubble from our dark matter particles) in a
    medium of density $\rho$ is
    \begin{eqnarray}
      F_d &=& \frac{c_d}{2}\rho v^2\pi R^2 
    \end{eqnarray}
    with $c_d$ being the drag coefficient of order unity, and so with a mass
    $M$ the velocity is diminished with the rate
    \begin{eqnarray}
      - \frac{dv}{dt}&=& \frac{F_d}{M}\\
      &=&  \frac{c_d\rho v^2\pi R^2}{2M}\\
      \hbox{giving } \qquad \frac{1}{v}&=& \int \frac{c_d\rho\pi R^2}{2M} dt\\
      &=& t  \frac{c_d\rho\pi R^2}{2M}- \frac{1}{v_{start}}.
    \end{eqnarray}
    Here $v_{start}$ is the velocity when the stopping of the
    pearl starts, say about 300 km/s for a pearl entering the Earth
    atmosphere and mountains. Formally 
    the pearl never stops, but order of magnitudewise the stopping
    takes place in the time $t =\frac{M}{\rho \pi R^2 v_{start}}$ (after
    putting $c_d$ and $2$ to be of order unity). So the distance of
    stopping becomes
    \begin{eqnarray}
      \hbox{``stopping length''} &=& t*v_{start}\\
      &=& \frac{M}{\rho \pi R^2 }.
    \end{eqnarray}
    
    Let us consider first the bubble inside a dust grain of radius 
    $0.1 \, \mu m$ satisfying the low velocity limit 
    $\frac{\sigma}{M}|_{v\rightarrow 0} = 15\, m^2/kg$, which has mass 
    $M = 3*10^{-15} \, kg$ (\ref{dgmfM}), radius
    $R= 1.2*10^{-9} \, m$ (\ref{dgmfR}) and 
    $\frac{\pi R^2}{M} = 1.5*10^{-3}\, m^2/kg$.  
    Taking for stone the density $\rho=3000kg/m^3$ we get
    \begin{eqnarray}
      \hbox{``stopping length''} &=& \frac{1}{1.5*10^{-3}m^2/kg*3000kg/m^3}
      = 0.22 m.
    \end{eqnarray}
   Similarly using the minimal size pearl which reaches down to 1400 m within
   one year, having mass $M = 6.5*10^{-14} \, kg$ (\ref{ftyfMc}), radius
   $R= 3.1*10^{-9} \, m$ (\ref{ftyfRc}) and 
   $\frac{\pi R^2}{M} = 4.6*10^{-4}\, m^2/kg$, we get
    \begin{eqnarray}
      \hbox{``stopping length''} &=& \frac{1}{4.6*10^{-4}m^2/kg*3000kg/m^3}
      = 0.72 m.
      \end{eqnarray}
    The corresponding stopping times are with $v =$ 300 km/s
    \begin{eqnarray}
       t &=& \frac{0.22m}{3*10^5m/s}= 7*10^{-7}s\label{st109}
       \end{eqnarray}
   and
     \begin{eqnarray} 
      t &=& \frac{0.72 m}{3*10^5m/s} = 2.4*10^{-6} s\label{st110}
      \end{eqnarray}
   respectively.
  \end{itemize}

  \begin{itemize}
  \item{\color{blue}Excitation}
    As long as the velocity is still over ca. 49km/s, collisions 
    with the nuclei in the
    shielding can excite the electrons inside the pearl by 3.5 kev or more
    and make pairs of quasi electrons and holes say. We expect that often the
    creation of (as well as the decay of) such excitations require electrons to
    pass through a (quantum) tunnel and that consequently
    there will be decay half lives of
    very different sizes. We hope even up to many hours or days or years...
  \end{itemize}

  \begin{itemize}
  \item{\color{blue} Slowly sinking:}
    After being stopped in the of the order of $\frac{1}{4}m$
    of the shielding
    the pearls continue with a much lower velocity driven by the gravitational
    attraction of the Earth.
    In section \ref{fty} we have discussed the minimal mass
    $M$ required to get the pearl come down in less than a year so that the
    seasonal variation in the counting by the DAMA-experiment can be
    realized; we found that the boundary number is not so far order of
    magnitudewise from our estimate based on assuming a size for the typical
    dust grain. Thus the time for the passage down of the pearl really tends
    to be not so far from about a year.
    After the year-long passage down the 1400 m  to the laboratories most of
    the pearls have returned to their
    ground states, but some exceptionally long living excitations may survive.
  \end{itemize}

  {\em Note that the slowly sinking velocity is so low that collisions with
    nuclei cannot give such nuclei enough speed to excite the
    scintillation counters neither in DAMA nor in Xenon-experiments} 

  \begin{itemize}
  \item{\color{blue} Electron or $\gamma$ emission}
    Typically the decay of an excitation could be that a hole in the Fermi sea
    of the electron cloud of the pearl gets filled by an outside electron
    under emission of another electron by an Auger-effect. The electron must
    tunnel into the pearl center. This can make the decay
    life time
    become very long and very different from case to case.
  That the decay energy is released most often as electron energy means that
  such events are
  discarded by most of the Xenon-experiments, which only expect the nucleus
  recoils to be dark matter events. 
  This would explain the long standing controversy
  consisting in DAMA seeing dark matter with a much bigger rate than the upper
  limits from the other experiments. 
\end{itemize}

  Using all the DAMA and DAMA-LIBRA data in the energy range 2 keV - 6 keV 
  \cite{DAMA2} gives
  an annual modulation amplitude of
  $0.0103 +/- 0.0008$ $cpd/kg/keV$ (cpd = counts per day).

  Rather recently though Xenon1T looked for potential excess events among the
  {\em electron recoil events} and found about  $16 \pm 5$ $events/year/tonne/keV$
  in the low energy region over a background of
  $(76 \pm 2)$ $events/year/tonne/keV$. This corresponds to a counting rate of
  about $0.00004$ $cpd/kg/keV$.
  In our model this rate should be compatible with the above DAMA-LIBRA event rate. However they deviate by a factor of about 250. 
  It therefore appears that we need
  the pearls to run much faster through the xenon apparatus than through the 
  DAMA one, as assumed in section \ref{fty}.
\section{3.5 keV}\label{threesame}

   Order of magnitudewise we see $3.5 keV$ in {\huge 3} different places.

  \includegraphics[scale=0.8]{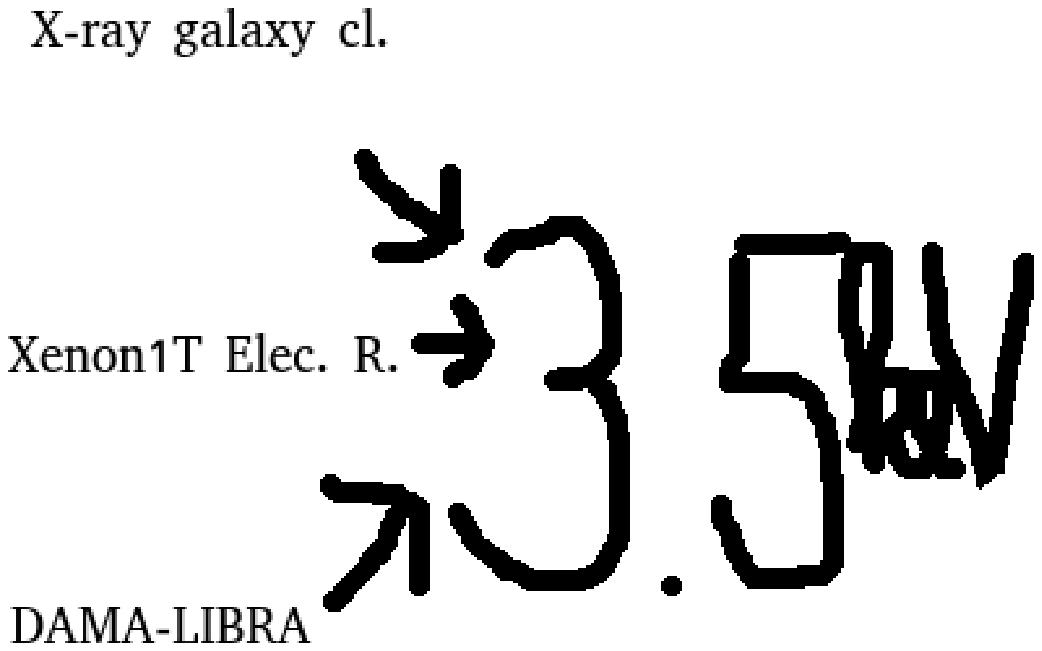}


  The energy level difference of about 3.5 keV occurring in 3 different
  places is important evidence motivating our 
  model of dark matter particles being excitable by 3.5 keV:
  \begin{itemize}
  \item{\color{blue} The line} From places in outer space with a lot of dark
    matter, galaxy clusters, Andromeda and the Milky Way Center, an unexpected
    X-ray line with photon energies of 3.5 keV (to be corrected for Hubble
    expansion...) was seen.
  \item{\color{blue} Xenon1T} The dark matter searching Xenon1T did not find
  the standard nucleus-recoil events expected for dark matter, but   
    found an excess of {\em electron-recoil} concentrated crudely around 3.5 keV.
  \item{\color{blue} DAMA} The seasonally varying component of their events lie
  in energy
    between 2 keV and 6 keV, not far from centering around 3.5 keV.
    \end{itemize}



  We take it seriously and not as an accident that both DAMA and Xenon1T see
  events with energies of the order of the controversial astronomical 3.5 keV X-ray
  line. We are thereby driven towards the hypothesis that the energies 
  for the events in these
  underground experiments for the events are determined from a decay
  of an excited particle, rather than from a collision with a particle in the
  scintillator material. It would namely be a pure accident, if a collision
  energy should just coincide with the dark matter excitation energy
  observed astronomically.

  {\em So we ought to have decays rather than collisions!}
  {\em How then can the dark matter particles get excited ?}
 
  You can think of the dark matter pearls in our model hitting electrons
  and/or nuclei on their way into the shielding:
  \begin{itemize}
  \item{\color{blue} Electrons} Electrons moving with the speed of the dark
    matter of the order of 300 km/s toward the pearls in the pearl frame
    will have kinetic energy of the order
    \begin{eqnarray}
      E_e &\approx & \frac{1}{2}*0.5 MeV*(\frac{300km/s}{3*10^5km/s})^2=
      0.25 eV.
    \end{eqnarray}
  \item{\color{blue}Nuclei} If nuclei are say Si, the energy in the
    collision
    will be 28*1900 times larger $\sim 5*10^4$ * $0.25 eV$
    $\approx$ 10 keV. That would allow a 3.5 keV excitation.

    To deliver such $\approx$ 10 keV energy the nucleus should hit
    something harder than just an electron inside the pearls. It should
    preferably hit a nucleus, e.g. C, inside the pearl. 
    \end{itemize}
  

\section{Direction of the Pull on Parameters from Fitting}

The fitting of our model is not quite perfect although taking into account
that it is so crude, what we assume, it really works reasonably well.

We can therefore seek to learn in what direction it should be improved
to get a better agreement, i.e. we could ask the direction of the pull
for our fitting.

The troubles really are:

\begin{itemize}
\item The combination $r_{dust}E$, where $E$ is the elastic modulus of the
  dust grain and $r_{dust}$ the size / radius of the dust pearl,    should be very
  large to make the value of the parameter $v_0$,
  which gives the characteristic velocity where the
  cross section to mass ratio $\frac{\sigma}{M}$ decreases significantly,
  agree with the value 
  $v_0 \approx$ 220 km/s suggested by fitting to the data of Figure \ref{Correa}.
\item We have some lower limits for the mass $M$ especially from the
  requirement that the pearls shall come sufficiently quickly through the
  shielding
  of the DAMA experiment,  so as to be able to deliver the seasonal variation
  being used by DAMA to identify them as dark matter. 
  This puts a constraint on $r_{dust}$ when we use the relation
  \begin{eqnarray}
    \frac{\sigma}{M}|_{v=0}&=&\frac{\pi r_{dust}^2}{M},
  \end{eqnarray}
  and the zero velocity limit $\frac{\sigma}{M}|_{v=0}$ obtained from the
  extrapolation of Correa's fit to the data in Figure \ref{Correa}. 
  \end{itemize}

Both of these constraints push the dust radius $r_{dust}$ upwards, but a
factor of $2*10^{11}$ increase in $r_{dust}$ from $10^{-7}m$ to $2*10^4m$, 
as would be required if we take the elastic modulus
$E$ to be just that for a normal material, 
would be too outrageous. However an increase in $r_{dust}$ is what is called for.

We could attempt to justify such an increase of $r_{dust}$ as an effect
of the hardening of the material, which would make it less easy to
split off pieces of the dust grain.

We also have the possibility to say, that the zero velocity cross section
to mass ratio is largely only an extrapolation and not a true measurement.
Thus we should rather say that we  should only trust the  combination
of $v_0$ and $\frac{\sigma}{M}|_{v=0}$ which gives us the value of the
ratio for the velocity region where it was truly measured by the
dwarf galaxies essentially.

Estimating crudely seeking to fit the formula (\ref{formula}) by
as small $v_0$ as possible we seemingly cannot get $v_0$ smaller than
about 100 km/s. And if we seek this type of fit the zero-velocity ratio
must be increased to about 1000 $cm^2/g$ rather than the used
150 $cm^2/g$. But this means that there is no chance for dramatically change
the value for $v_0$ which we need.

It is thus hard to see how we can get a sensible fit to the velocity dependence
of the cross section without the unbelievably large modulus, or very big dust
grains.

\section{Conclusion}



  	We have described a seemingly very viable model for dark
    matter consisting of
    nano-meter size but macroscopic pearls.
    These pearls consist of a
    bubble of a new speculated type of vacuum containing some 
    material - presumably carbon - under the high pressure of the skin
    (surface tension).
    It can contain about
    $10^{12}$
    nucleons in the
    bubble of radius about
    a few nanometers.
    
  	The electrons in a pearl are partly pushed out of the genuine
    bubble of the new vacuum phase.

	Further outside the bubble with its surrounding electron cloud
  there is an a priori usual dust grain, but it is modified somewhat
  in its density and strength - especially the elastic modulus -
  by the modification of the electron wave functions due to the for usual
  matter exceedingly big homolumo gap.



  We have compared the model or attempted to fit:
  
  \begin{itemize}
  \item { Astronomical suggestions for the self interaction of dark matter
    in addition to pure gravity}
  \item { The astronomical 3.5 keV X-ray emission line found by satellites, 
  	supposedly from dark matter.}
    \item { The underground dark matter searches.}
    
    \end{itemize}
    

  We list below the quantities we have crudely estimated:
  \begin{enumerate}
  \item
  The low velocity cross section divided by mass is not
    predicted in our present model, but the 
    pearl mass estimated using the value Correa obtained \cite{CAC} from
    the dwarf galaxy data is only in very mild conflict with the mass
    constraints in our model - see Table \ref{Bounds}
    in section \ref{bounds} below.
  \item 
  That the signal from Xenon1T and Dama should agree
    except for different scintillation efficiencies and notation, ...
    ({\em not working so well!}). However, we suppose it is solved by the
    pearls having higher terminal velocities in the {\em fluid} xenon, than
    in the solid NaI in DAMA.
  \item
  The absolute rate of the two underground experiments.
  \item 
  The rate of emission of the 3.5 keV line from the Tycho supernova remnant
  \cite{Jeltema} due to the excitation of our pearls by cosmic rays \cite{Bled20}.
  \item 
  Relation between the frequency 3.5 keV and the overall
  emission rate of this X-ray line observed from galaxy clusters etc.
  \item 
  Using a ``dirty'' story of hardening the dust grain,
    we can get the velocity scale parameter $v_0$ in our velocity dependence
    fit of the $\frac{\sigma}{M}$ data in Figure \ref{Correa}.
  \item 
  We have previously predicted the ratio of dark matter to ordinary matter
  in the Universe to be of the order of 5 by consideration of the binding 
  energies per nucleon in helium and heavier nuclei, assuming that the 
  ordinary matter at some time about 1 s after the Big Bang was expelled 
  from the pearls under a fusion explosion from He fusing into say C 
  \cite{Dark1}.
    \end{enumerate}

  \subsection{R\'esum\'e of our Predictions}
  \label{predictions}


  
\begin{center}
  \begin{table}
  \begin{tabular}{|c|c|c|c|c|c|}
  \hline
  \# \& exp/th & Quantity &value  &related Q. & value&sec.\\
  \hline
  1.& Dwarf Galaxies&&&\\
  exp &Velocity par. $v_0$ & $220km/s$
  & $4r_{dust}E$ &
  $8.1*10^{13}kg/s^2$&\ref{vdp}\\
  th. & with hardening &$77 km/s$&$4r_{dust}E$    &$1*10^{13}kg/s^2$&\\
  th. & without hard.& $0.7 cm/s$&$4r_{dust}E$&$400 kg/s^2$&\\
  \hline
  2.& DAMA-LIBRA&&&&\ref{absrates}\\
  exp &
  &$0.041 cpd/kg$ &suppression
  &
  $1.6*10^{-10}$&\\
  th& air
  &$
  0.16 cpd/kg$&&$
  6*10^{-10}$
  &\\
  th&stone & $1.6*10^{-5}cpd/kg$ &&$6*10^{-14}$&\\
  \hline
  3.&Xenon1T&&&&\ref{absrates}\\
  exp&
  &$
  2*10^{-4} cpd/kg$&suppression
  &$
  6*10^{-13}$&\\
  th & air &$
  0.16 cpd/kg$&&$
  6*10^{-10}$\\
  th& stone &$1.6*10^{-5}cpd/kg$&&$6*10^{-14}$&\\
  \hline
  4.&Jeltema \& P.&&&&\ref{Jelt}\\
  exp&counting rate&$2.2 *10^{-5}phs/cm^2/s$&$\frac{\sigma}{M}|_{Tycho}$
  &$5.6*10^{-3}cm^2/kg$&\\
  th&&$3*10^{-6}phs/s/cm^2$&$1\% *\alpha*\frac{\sigma}{M}|_{nuclear}$
  &$8*10^{-4}cm^2/kg$&\\
  \hline
  5.&Intensity 3.5 kev&&&&\ref{Xray}\\
  exp& $\frac{N\sigma}{M^2}$ & $10^{23}cm^2/kg^2$
  &$\frac{\xi_{fS}^{1/4}}{\Delta V}$&$0.6 MeV^{-1}$&\\
  th& & $3.6*10^{22}cm^2/kg^2$&&$0.5 MeV^{-1}$&\\
  \hline
  6.& Three Energies&&&&\ref{threesame}\\
  ast& line & 3. 5 keV&&&\\
  DAMA& av. en.& 3.4 keV&&&\\
  Xen.&av. en. &3.7 keV&&&\\
  \hline
\end{tabular}
\caption{\label{Predictions}}
\end{table} 
\end{center}


In Table \ref{Predictions} we summarize six of our predictions.
In the first column is written the number in our listing and we use the symbols
''exp" = experiment meaning the number given in the third column 
corresponds to the experimentally observed number or is closely related to it,
and ''th" = theory meaning that the third column in that row contains our model
prediction. The second column contains a short name for the quantity we attempt
to predict plus in some cases a detail of the assumption in our model, such
as ``air'' or ``stone'' specifying the material in which the dark matter
particle is supposed to be stopped. In most cases we formulate the calculation
as to be found in the section listed in the sixth column by calculating
a quantity, related to the value measured and given in column 2, called
``related quantity'' = ``related Q.'' and specified in column 4, by its name or
formula. The experimental and theoretical values of this related 
quantity ``related Q.'' are given in the appropriate ''exp" and ''th" lines of
column 5, with the possible detailed choice of the version 
of our model spelled out by the words in column 2 like ''air" or ''stone".
In the second column of item 6 we use the shorthand ''av. en." to denote our 
best estimate of the average energy of the supposedly dark matter caused events.


Let us now briefly explain the 6 predictions in Table \ref{Predictions}:

\begin{enumerate}
\item{{\color{blue} Dwarf Galaxies, Velocity parameter $v_0$} } We fit the
  velocity dependence as found by Correa \cite{CAC} of the cross section to
  mass ratio $\frac{\sigma}{M}$
  to a formula for relatively small velocities of the form
  \begin{eqnarray}
    \frac{\sigma}{M} &=& \frac{\sigma}{M}|_{v\rightarrow 0}*(1-v/v_0)^4.
  \end{eqnarray}
  The velocity parameter $v_0$ turned out to depend on the
  ``related Q.'' being $4r_{dust}E$, where $r_{dust}$ is the radius of the
  dust particle around the bubble and $E$ is the elastic modulus of this
  dust grain. In column 2 we distinguish two versions of our theoretical
  model: ``with hardening'' meaning that we use an effect which the
  very strong homolumo gap in the bubble is supposed to have on the dust
  around it. It makes the modulus $E$ much bigger and we get the
  quantity $4r_{dust}E$ raised up to $1*10^{13}kg/s^2$ 
  as written in column 5.
  Assuming no such effect we have the version of our model ``without hard.''
  described in the third line of item 1.
\item{{\color{blue} DAMA-LIBRA }} The counting of the seasonal
  varying component of events in the DAMA-LIBRA experiment is given as
  number of counts per day ``cpd'' and taken per kg of scintillator material in
  which it is seen, so that the seasonal varying counting rate $S_m$
  (m for modulation) is given as $0.041cpd/kg$. We then predict this
  rate by an extremely crude argument first based on assuming that the energy
  from the kinetic energy of the incoming dark matter particles
  is equally distributed over all the matter in the Earth 
  down to the 1400 m depth of the experiments. But we also expect a further
  suppression factor called $suppression$ in column 4 arising from the fact that 
  the energy observed by DAMA-LIBRA comes from decays of excitations of the dark
  matter pearls having exceptionally long living excitation states
  with lifetimes of the order of the passage time down to the experimental halls.
  For the fraction $suppression$ of the kinetic energy surviving down to the
  depth of the experiment we take, by a dimensional argument, the ratio
  of the time of excitation (in air or in stone) relative to the passage time
  down to the experiment. The theoretical numbers $6*10^{-10}$ for air and
  $6*10^{-14}$ for stone are simply such ratios of 
  excitation times to passage time.
\item{{\color{blue} Xenon1T}} This is just the same estimation of the
  rate from crude dimensional arguments as in the DAMA-LIBRA case. Here though
  one has no modulation involved. Rather the experimental numbers are
  extracted as the observed electron recoil events minus an
  estimated background.
\item{{\color{blue} Jeltema \& P.}} Jeltema and Profumo\cite{Jeltema} observed
  the 3.5 keV X-ray line supposedly coming from dark matter from the
  supernova remnants of the Tycho Brahe supernova. We suppose this to be due
  to 3.5 keV radiation from the relatively small amount of dark matter in the
  neighbourhood of the remnant being, however, excited by a very high intensity
  of cosmic rays, for which the dark matter shows an effective
  $\frac{\sigma}{M}$ as if it were just atomic nuclei. The fraction of
  order the fine structure constant $\alpha$ of the excitation energy is
  emitted as 3.5 keV X-rays, while the rest of the emission is dominated by
  electrons. The $1\%$ in the fourth column alludes to the
  fraction of the supernova energy coming as cosmic rays, while
  the ratio $\frac{\sigma}{M}_{nuclear}$ is the above mentioned ratio
  taken assuming that, as far as the cosmic ray cross section is concerned, 
  the dark matter pearl 
  just behaves like a group of independent nuclei of too small a size to
  significantly shadow  each other. In the experimental row
  we find the effective $\frac{\sigma}{M}$ that would provide the
  intensity of the 3.5 keV line observed by Jeltema and Profumo
  $2.2*10^{-5}phs/cm^2/s$ (where phs means photons), it is called
  $\frac{\sigma}{M}|_{Thyco}$.
\item{{\color{blue} Intensity}} We fitted both the overall intensity of the
  3.5 keV X-ray line observed from galaxy clusters etc.
  as supposedly proportional to the
  local {\em square} of the dark matter density and 
  the very 3.5 keV energy itself identified with
  the homolumo gap in the electron spectrum of the dark matter pearls.
  These two quantities depend on the 
  same combination of parameters in our model $\frac{\xi_{fS}^{1/4}}{\Delta V}$
  for which we found the values in the 5th column. 
  Formally we here take the 3.5 keV
  homolumo gap used not as an experimental measurement, but just
  as a parameter used in the theory. But oppositely we somewhat arbitrarily
  considered the intensity measurement a genuine experiment formally.
  
\item{{\color{blue} Three Energies}} We here just emphasize the very important
  observation for our theory, that the energy number 3.5 keV per event pops
  up three times. Instead of formally considering one of the numbers as
  theory and another as experiment, we simply list in column 1 the name of the
  experiment giving the energy per event then listed in column 3.
  
\end{enumerate}


\subsection{Bounds and Estimates of Size: Mass or Radius} \label{bounds}

\begin{table}
\begin{tabular}{|c|c|c|c|c|c|c|}
  \hline
  Description&$R$&$\Delta R$&$M$&$\Delta M$&Sec.& formula\\
  \hline
  $\frac{\sigma}{M}|_{v\Rightarrow 0}=15\frac{m^2}{kg}$&$1.2*10^{-9}m$&70\%
  &$\approx 3*10^{-15}kg $ &200\%&\ref{dgm}&(\ref{dgmfM}),
  (\ref{dgmfR})\\
  \hline
  Faster than year&$\ge 1.0*10^{-9}m$&&$\ge 2.1*10^{-15}kg$&
  &\ref{fty}&(\ref{ftyfR}),(\ref{ftyfM})\\
  Corrected f.t.y.& $\ge 3.1*10^{-9}m$&&$\ge 6.5*10^{-14}kg$&
  &\ref{fty}&(\ref{ftyfRc}),(\ref{ftyfMc})\\
  \hline
  Dust enough&$\ge 1.0*10^{-9}m$&&$\ge 2*10^{-15}kg$&&\ref{ndg}
  &(\ref{ndgfR}), (\ref{ndgfM})\\
  \hline
  Velocity dep.& $\approx 10^{-8}m$&big&$\approx 10^{-13}kg$
  &big& \ref{vdp}&\\
  w. E= $400^4$&$10^{-10}m$&&$\approx 2*10^{-18}kg$&&\\
  \hline
   \end{tabular}
  \caption{\label{Bounds}}
  \end{table}

In Table \ref{Bounds} we summarize the various estimates and bounds on 
the mass and radius of our pearls, giving a reference to the relevant section
and formula in each case. 

The uncertainties given in this table are only extremely crude and supposed
to be interpreted logarithmically, so that 200\% means within about a factor 7
($=e^2$) and 70\% within a factor of order 2. The ``big'' means that the
uncertainty is huge because the mass essentially comes in only via its
sixth root $\sqrt[6]{M}$.

\end{document}